\documentclass{aa} 
\usepackage{textcomp, gensymb}
\usepackage{caption}

\DeclareCaptionFormat{myformat}{\textbf{#1#2}#3}
\usepackage{graphicx}
 \usepackage[varg]{txfonts}
\usepackage{color}
\definecolor{VividBlue}{RGB}{0, 163, 224}
\usepackage[colorlinks=true,citecolor=VividBlue,urlcolor=blue,linkcolor=black]{hyperref}
\hypersetup{colorlinks=true, urlcolor=VividBlue}

\usepackage{booktabs}
\usepackage{subcaption} 

\graphicspath{{./figs/}}
\newcommand{\xmmnewton}{\textsl{XMM-Newton}\xspace}

\usepackage{footmisc}
\usepackage[dvipsnames]{xcolor}
\usepackage{float}

\makeatletter
\def\@makefnmark{\hbox{\@textsuperscript{\normalfont\color{red}\@thefnmark}}}
\makeatother

\DeclareUnicodeCharacter{2212}{-}

\begin{document}

\title{
Variable structures in the stellar wind of the HMXB Vela X-1
}

\author{%
L.~Abalo\inst{\ref{affil:cosine},\ref{affil:Leiden}} \and
P.~Kretschmar\inst{\ref{affil:ESAC}}\and
F.~F\"urst\inst{\ref{affil:ESAC}}\and
C.~M.~Diez\inst{\ref{affil:ESAC}}\and
I.~El~Mellah\inst{\ref{affil:FisicaUSACH},\ref{affil:CIRAS}} \and
V.~Grinberg \inst{\ref{affil:ESTEC}}\and
M.~Guainazzi\inst{\ref{affil:ESTEC}}\and
S.~Mart\'inez-N\'u\~nez\inst{\ref{affil:IFCA}}\and
A.~Manousakis\inst{\ref{affil:Sharjah}}\and
R.~Amato\inst{\ref{affil:INAF-OAR}} \and
M.~Zhou\inst{\ref{affil:IAAT}} \and
M.W. Beijersbergen\inst{\ref{affil:cosine},\ref{affil:Leiden}} 
}

\institute{%
cosine measurement systems, Warmonderweg 14, 2171 AH  Sassenheim, The Netherlands \label{affil:cosine}
\email{l.abalo@cosine.nl}
\and 
Huygens-Kamerlingh Onnes Laboratory, Leiden University, Postbus 9504, 2300 RA Leiden, The Netherlands \label{affil:Leiden}
\and  
European Space Agency (ESA), European Space Astronomy Centre (ESAC), Camino Bajo del Castillo s/n, 28692 Villanueva de la Cañada, Madrid, Spain \label{affil:ESAC}
\and 
Departamento de F\'isica, Universidad de Santiago de Chile, Av. Victor Jara 3659, Santiago, Chile \label{affil:FisicaUSACH}
\and 
Center for Interdisciplinary Research in Astrophysics and Space Exploration (CIRAS), USACH, Chile \label{affil:CIRAS}
\and 
European Space Agency (ESA), European Space Research and Technology Centre (ESTEC), Keplerlaan 1, 2201 AZ Noordwijk, The Netherlands \label{affil:ESTEC}
\and 
Instituto de F\'isica de Cantabria (CSIC-Universidad de Cantabria), E-39005, Santander, Spain \label{affil:IFCA}
\and 
Department of Applied Physics and Astronomy, College of Sciences and Sharjah Academy for Astronomy Space Sciences, and Technology (SAASST), University of Sharjah, POBox 27272 Sharjah, UAE \label{affil:Sharjah}
\and 
INAF -- Osservatorio Astronomico di Roma, Via Frascati 33, I-00040, Monte Porzio Catone (RM), Italy \label{affil:INAF-OAR}
\and 
Institut f\"ur Astronomie und Astrophysik, Universität T\"ubingen, Sand 1, 72076 T\"ubingen, Germany \label{affil:IAAT}
}

\date{Received: 28 March 2024 / Accepted: 12 October 2024}

\abstract
{
Strong stellar winds are an important feature in wind-accreting high-mass X-ray binary (HMXB) systems. Exploring their structure provides valuable insights into stellar evolution and their influence on surrounding environments. However, the long-term evolution and temporal variability of these wind structures are not fully understood.   
}
{
This work probes the archetypal wind-accreting HMXB \object{Vela X-1} using the Monitor of All-sky X-ray Image (MAXI) instrument to study the orbit-to-orbit absorption variability in the $2–10$~keV energy band across more than 14 years of observations. Additionally, the relationship between hardness ratio trends in different binary orbits and the spin state of the neutron star is investigated. 
}
{
We calculate X-ray hardness ratios to track absorption variability, comparing flux changes across various energy bands, as the effect of absorption on the flux is energy-dependent. We assess variability by comparing the hardness ratio trends in our sample of binary orbits to the long-term averaged hardness ratio evolution derived from all available MAXI data. 
}
{
Consistent with prior research, the long-term averaged hardness ratio evolution shows a stable pattern. However, the examination of individual binary orbits reveals a different hardness ratio evolution between consecutive orbits with no evident periodicity within the observed time span. We find that less than half of the inspected binary orbits align with the long-term averaged hardness evolution. Moreover, neutron star spin-up episodes exhibit harder-than-average hardness trends compared to spin-down episodes, although their distributions overlap considerably.
}
{
The long-term averaged hardness ratio dispersion and evolution are consistent with absorption column densities reported in literature from short observations, indicating that a heterogeneous wind structure -- from accretion wakes to individual wind clumps -- likely drives these variations.
The variability observed from orbit to orbit suggests that pointed X-ray observations provide limited insights into the overall behaviour of the wind structure. 
Furthermore, the link between the spin state of the neutron star and the variability in orbit-to-orbit hardness trends highlights the impact of accretion processes on absorption. This connection suggests varying accretion states influenced by fluctuations in stellar wind density.
}

\keywords{X-rays: individuals: Vela X-1 -- X-rays: binaries -- Stars: winds, outflows -- Accretion, accretion disks
}

\maketitle

\section{Introduction}
\label{section:introduction}

\begin{figure} 
\centering
\includegraphics[width=0.49\textwidth]{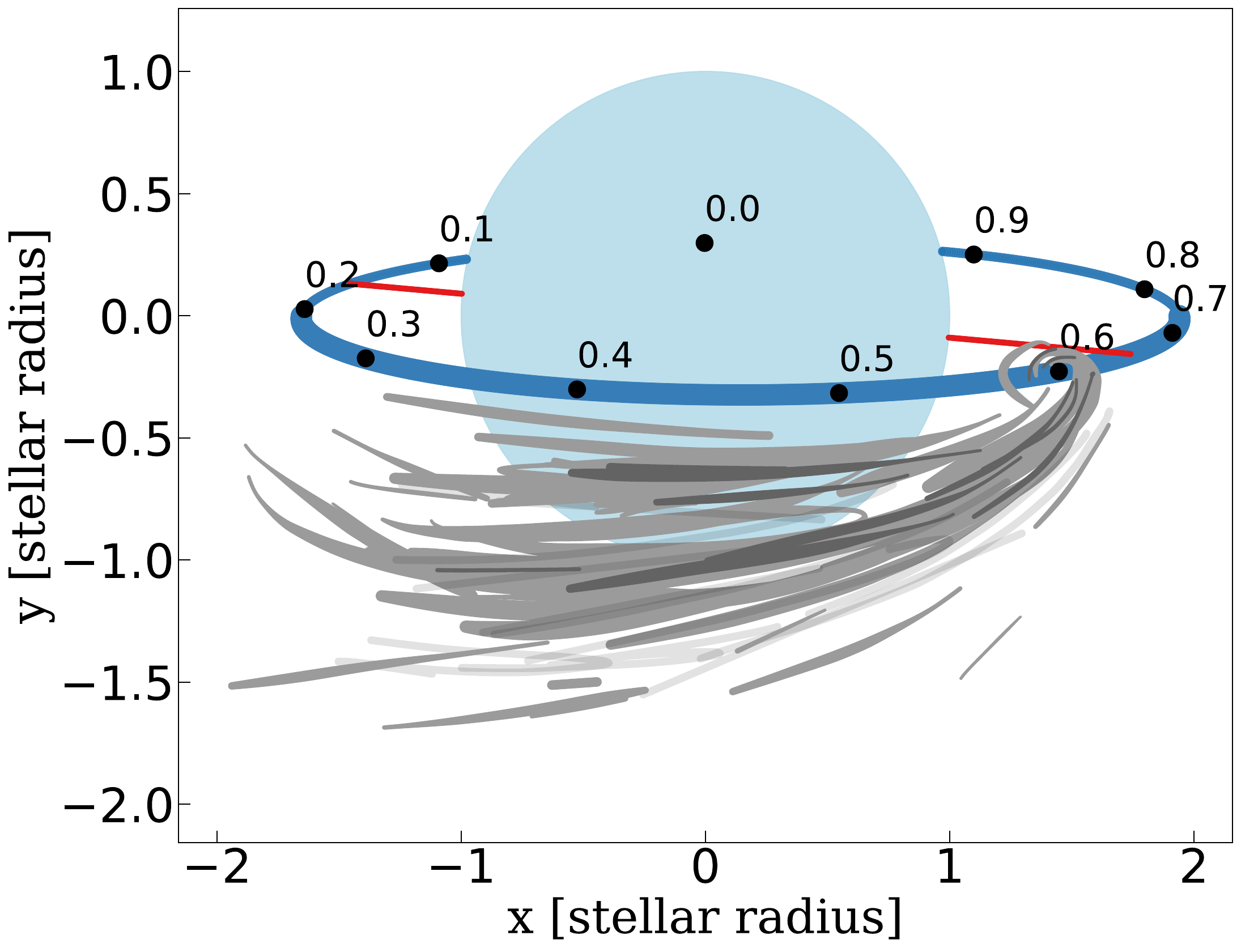}
\caption{
Vela X-1 as seen from Earth, with an orbital inclination $i>73\degree$ compatible with the known constraints \citep[with parameters from][]{Kreykenbohm_2008}. The black points represent the positions of the neutron star at specific phases (with origin at mid-eclipse time). The red line is the main axis of the eccentric orbit. Owing to the eccentricity, inferior conjunction of the neutron star occurs between $\phi_{\mathrm{orb}}=0.4$ and $\phi_{\mathrm{orb}}=0.5$. The accretion wake trailing the neutron star is represented at a fiducial orbital phase $\phi_{\mathrm{orb}}=0.6$.
}
\label{figure:phases}
\end{figure}

\object{Vela X-1} (\object{4U 0900$-$40}) is among the best-studied X-ray sources since its discovery in 1967 \citep{Chodil_1967}. 
The system is an eclipsing high-mass X-ray binary (HMXB) composed of a B0.5 Ia donor star, HD 77581  \citep{Hiltner_1972}, and an accreting neutron star with a pulse period of $\sim$283 s \citep{McClintock_1976} in a low eccentric orbit ($e\sim 0.0898$) of $\sim$8.964~days \citep{Kreykenbohm_2008}. The high inclination of $>73\degree$ \citep{vanKerkwijk_1995} leads to regular X-ray eclipses of the neutron star, resulting in reduced observed X-ray emission, particularly in the hard X-ray band, for nearly 20\% of the orbit. The neutron star has a magnetic field of $\sim$2.6$\times10^{12}$ G \citep{Fuerst_2014a}. The close distance of the system, $1.99^{+0.13}_{-0.11}$ kpc \citep{Kretschmar_2021a}, makes it one of the brightest X-ray objects in the sky despite its modest mean luminosity of $5\times10^{36}$ erg s$^{-1}$ \citep{Fuerst_2010a}. 
The neutron star has an estimated mass of $\sim1.7-2.1~M_\odot$ \citep{Kretschmar_2021a} and the radius of HD 77581 is 30 $R_{\odot}$ \citep{vanKerkwijk_1995}. The mean orbital separation of $\sim$1.7 $R_{\star}$ \citep{Quaintrell_2003} causes the neutron star to be heavily embedded in the dense and line-driven wind of the companion, which has a mass loss rate of $\sim$10$^{−6}\,M_{\odot}\,\mathrm{yr}^{−1}$ \citep{Gimenez-Garcia_2016}. 
A fraction of the wind material is accreted and channelled to the magnetic poles of the neutron star, feeding the X-ray emission. In Fig.~\ref{figure:phases}, we represent the orbit of the neutron star around HD 77581 as seen from Earth.
For a comprehensive overview, we refer to \citet{Kretschmar_2021a}. 

HMXBs like Vela~X-1 offer the opportunity to investigate both accretion onto compact objects, and the structure of winds from massive stars. To characterise the latter, we can use the neutron star as an X-ray probe orbiting in the wind and analyse the variability of the X-ray absorption along the line-of-sight. 
As commonly done in X-ray astronomy, we will give the absorbing column $N_\textrm{H}$ in units of hydrogen atoms per cm$^{2}$, where we assume a certain abundance of the heavier elements like O, S, Si, and Fe relative to H. This abundance vector is based on the solar abundance given by \citet{Wilms_2000}. That is, the column is given as an equivalent hydrogen column, even though hydrogen does not contribute significantly to the absorption in the X-ray band. 
Understanding the structure and environmental impact of winds is important for studies of massive star evolution\citep{Martinez_2017}. The strong stellar winds produced by the donor stars are driven by line scattering of the star’s intense continuum radiation field \citep{Lucy_Solomon_1970,CAK_1975}. 
Line-driven winds are unstable, with the line-deshadowing instability believed to be responsible for the formation of overdense inhomogeneities or “clumps” within a more diffuse plasma environment \citep{Owocki_1984, Owocki_1988, Feldmeier_1995}. 
In addition, large-scale structures are expected to form in the stellar wind by the presence of the orbiting X-ray source influencing the wind flow by its gravity and the impact of the X-rays on the wind acceleration as further detailed below.

The accretion wake arises as a result of the wind beaming induced by the gravitational field of the neutron star. This gravitational pull causes the stellar wind to be focused towards the neutron star, forming a concentrated flow that generates an unsteady bow shock in the vicinity of the neutron star \citep{Blondin_1991, Manousakis_Walter_2015a, Malacaria_2016a}. The accretion wake is characterised by dynamic and evolving structures as the stellar wind interacts with the neutron star's gravity. 
The photoionization wake is generated by the X-ray radiation emitted by the neutron star. 
Similar to the concept of the Str\"omgen sphere around the blue star, iso-$\xi$ surfaces are defined based on the ionisation parameter $\xi$ around the neutron star \citep[for more details, see e.g.,][]{Hatchett_1977, Manousakis_Walter_2015a, Manousakis_Walter_2015b, Sander_2018}. 
The X-ray radiation ionises the material within this region, forming a wake-like structure downstream of the neutron star. This photoionization wake significantly influences the ionisation state and physical properties of the nearby gas \citep[see, e.g.,][]{Amato_2021}.  

The intrinsic X-ray flux of Vela X-1 varies on a wide range of time scales. 
Similarly to other X-ray pulsars, fluctuations within a pulse period are commonly observed \citep{Staubert:1980, Kretschmar_2014}. 
Vela X-1 also exhibits bright flares \citep{Kreykenbohm_2008, Martinez_2014, Lomaeva_2020a} and "off-states" on longer time sales \citep{Inoue_1984, Kreykenbohm_1999a, Kreykenbohm_2008, Doroshenko_2011a, Sidoli_2015a}; 
in both cases, flux can quickly change by more than an order of magnitude. 
These variations are usually explained by enhanced accretion mediated by instabilities at the outer rim of the neutron star magnetosphere. 
Overall, the flux variations have been found to follow a log-normal distribution \citep{Fuerst_2010a}. 

Similarly to other accreting X-ray pulsars, the X-ray spectrum of Vela X-1 outside eclipse is characteristically described by a power-law with a turnover or cutoff at high energies, typically beyond 15-30 keV. This spectral profile is further modified by frequently strong absorption, an occasional added soft component, a prominent iron line, and cyclotron resonance scattering features in the hard X-rays \citep[e.g.][]{Haberl_White_1990, Kreykenbohm_1999a, Odaka_2013, Fuerst_2014a, Diez_2022, Diez_2023}. The photon index of the power-law component exhibits mild variability, generally within a few percent, and shows a trend of spectral hardening with increasing brightness \citep{Odaka_2013, Fuerst_2014a}.
In some studies \citep{Martinez_2014, Malacaria_2016a}, the power-law index is constant when a complex absorption component is included. However, in studies that allow for variations in the power-law index, changes in spectral hardness are mostly attributed to changes in the derived absorption \citep{Haberl_White_1990, Diez_2023}. 

The origin of this variable absorption is likely twofold. 
Given the high inclination of Vela X-1, the accretion wake is expected to intercept the line-of-sight as the neutron star orbits its stellar companion. 
The presence of the accretion wake from one orbit to another leads to a pattern in the orbital profile of $N_{\mathrm{H}}$ where absorption is generally higher (lower) when the neutron star moves away from us (towards us), in agreement with the predictions from \citet{Manousakis_Walter_2015b}. 
However, the absorption derived at the same orbital phase can be very different in observations taken during different orbits, indicating variable absorbing structures between individual binary orbits \citep[][ see also Fig.~\ref{figure:nh}]{Kretschmar_2021a}. 
This effect is generally ascribed to unaccreted clumps in the wind which serendipitously intercept the line-of-sight, leading to an increase in $N_{\mathrm{H}}$ on timescales of minutes to hours \citep{Grinberg_2017, El_Mellah_2020a, Diez_2022}. 
In order to separate these two absorbing components, we need to monitor absorption variations over many orbits. The Monitor of All-sky X-ray Image (MAXI) instrument is particularly well-suited for this task due to its continuous coverage of the entire orbit, enabling the study of orbit-to-orbit absorption variability that pointed X-ray observations, which cover only a small fraction of a single orbit, cannot provide. 

\begin{figure} 
\centering
\includegraphics[width=0.49\textwidth]{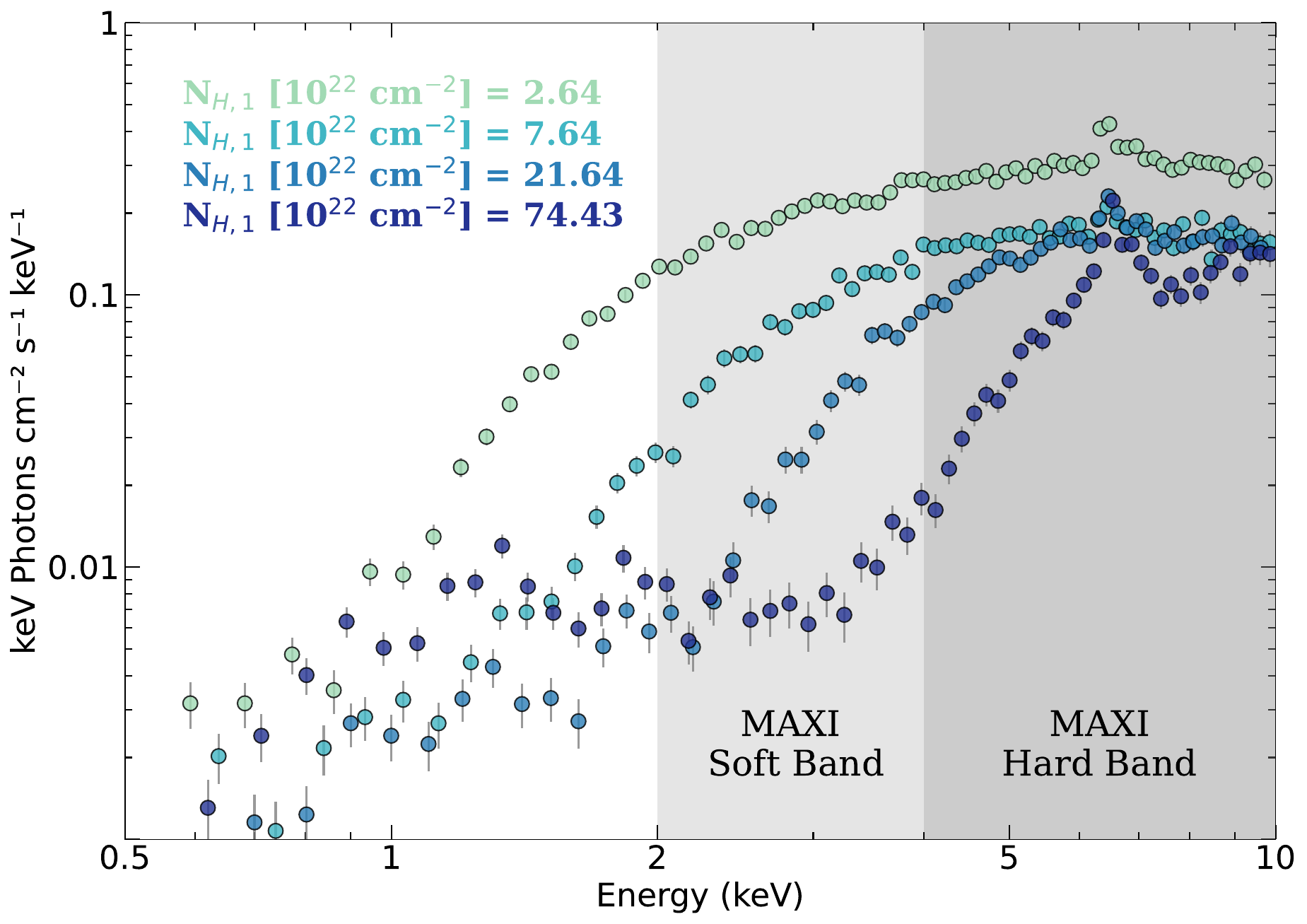}
\caption{
Four unfolded spectra of Vela X-1 from \xmmnewton observation (2019) taken with EPIC-pn. Each spectrum is $\sim$283$\,\mathrm{s}$ and corresponds to a different absorption column density value, from low $N_{\mathrm{H}}$ (light green) to high $N_{\mathrm{H}}$ (dark blue). The observation covers the 0.37--0.51 orbital phase range \citep{Diez_2023}. The shaded zones cover the chosen MAXI/GSC energy bands (2--4 keV and 4--10 keV) for our analysis.
}
\label{figure:xmm}
\end{figure}

The spin period of the neutron star in Vela X-1 has been monitored over decades, revealing periods of spinning up and down that can last up to months. These variations may be linked to episodes of enhanced accretion. This is supported by studies such as that of OAO 1657--415, where fluorescence lines suggest that stochastic increases in absorption might be associated with enhanced accretion episodes \citep{Pragati_2019}. Changes in the spin state of the neutron star have been connected to the accreted material, though the exact mechanisms are not fully understood \citep{Bozzo_2008}. Recent findings by \citet{Liao_2022} report a positive correlation between spin-up events and absorbing column density, attributing this to episodes of increased mass loss from the donor star. They propose that spin-up and spin-down periods correspond to prograde and retrograde orbits of the accretion disk, respectively. However, this scenario is challenged by the absence of observed global variations in line-driven winds from massive stars, except in co-rotating interaction regions \citep{Lobel_2008}. Additionally, the traditional model of accretion-induced torques, which involves coupling between a truncated accretion disk and a neutron star's magnetosphere \citep{Ghosh_1979}, is questioned in the context of Vela X-1 \citep{Kretschmar_2021a}. Alternative mechanisms that produce both positive and negative torques based on the angular momentum of the accretion flow have been suggested \citep{Shakura_2012}.

In this study, we use MAXI to continuously monitor Vela X-1 to probe the variable structures in the stellar wind of the system. In particular, we focus on assessing the orbit-to-orbit absorption variability and investigate the spin state connection. Our research enhances studies conducted with pointed X-ray observations, which only cover a fraction of a single orbit. 
It should be noted that MAXI data allow for the construction of detailed spectra only when integrated over long time periods, as demonstrated in \citet{Malacaria_2016a} and \citet{Liao_2022}. Therefore, only hardness ratios are discussed.
We demonstrate the impact of varying absorption in Fig.~\ref{figure:xmm}, which displays four different X-ray spectra taken by \xmmnewton, along with the corresponding absorbing column densities derived by \citet{Diez_2023}. Periods of high absorption ($N_{\mathrm{H}}\gtrsim 20\times 10^{22}\,\mathrm{cm^{-2}}$) lead to a substantial reduction in flux, particularly in the 2--4 keV band. These periods also reveal strong and ionised emission lines that can dominate the spectrum at lower energies. However, these lines mainly appear outside the energy range covered by the MAXI instrument employed in this work \citep{Watanabe_2006}.
Moreover, the methodology presented in this work encourages further investigations into similar HMXB systems. 
The instrument and dataset are described in Sect.~\ref{section:MAXI}. Our methods and data analysis are explained in Sections~\ref{section:overall_data_distribution_analysis} and \ref{section:orbital_profiles_analysis}. Finally, we discuss our results in Sect.~\ref{section:discussion} and provide a comprehensive summary of our findings and outline future prospects in Sect.~\ref{section:conclusion}.  

\section{MAXI observations}
\label{section:MAXI}

\subsection{MAXI}
\label{subsection:dataset}

Vela X-1 is regularly monitored through MAXI \citep{Matsuoka_2009}. This instrument is positioned on the International Space Station (ISS) within the Japanese Experiment Module and initiated its nominal observations in August 2009. It incorporates two types of X-ray slit cameras working in complementary ways: the Solid-State Slit Camera \citep[SSC,][]{Tomida_2011} operating within the energy range of 0.5--12 keV, and the Gas Slit Camera \citep[GSC,][]{Mihara_2011} in the 2--30 keV energy band. \par

We employ the MAXI/GSC in our study, which consists of six slit camera units, each housing two Xe-gas proportional counters equipped with one-dimensional slit-slat collimators. The field of view is 1.5$^\circ\times160^\circ$, limited by the ISS structure and solar paddle. The twelve gas counters achieve a total detector area of 5350 cm$^\mathrm{2}$. The instrument has an all-sky coverage of $\sim$85$\%$ per ISS orbit and $\sim$95$\%$ per day. It scans a point source from 40 to 150 s, contingent upon the source incident angle \citet{Sugizaki_2011}.  

\subsection{Dataset}
\label{subsection:dataset}

We use the ISS orbit light curve data\footnote{MAXI dataset employed for the analysis available at: \url{http://maxi.riken.jp/star_data/J0902-405/J0902-405.html}} to access the most detailed temporal information available over more than 14 years of MAXI observations as provided by the MAXI team (see Table~\ref{table:1}).  

\begin{table} 
\caption{Summary of MAXI dataset of Vela X-1.}
\begin{center}
\begin{tabular}{cc}
\hline
\hline
Parameter & Value \\ 
\hline
$\Delta T_{\mathrm{Obs}}$ & 55055 MJD -- 60236 MJD \\
& 2009, August -- 2023, October \\
\hline
Exposure time$^{\mathrm{(a)}}$ & $\sim$2.28 Ms \\
\hline
$\#$ Binary orbits$^{\mathrm{(b)}}$ & 579 \\
\hline
\end{tabular}
\end{center}
\tablefoot{$^{\mathrm{(a)}}$ Calculated based on source visibility of 40-150 seconds for each scan cycle \citep{Sugizaki_2011}, and taking into account the number of scans. It is provided as an estimate for the reference of the reader; 
$^{\mathrm{(b)}}$ Some binary orbits may be empty due to non-uniform data coverage throughout the observation period.
}
\label{table:1}
\end{table}

The light curve data are available in three energy bands: 2--4~keV, 4--10~keV and 10--20~keV. The source fluxes provided assumed photon fluxes calculated under the assumption of a Crab-like spectrum (count rates), where the 2--20~keV value typically differs from the sum of the three individual bands. For simplicity, we henceforth use the term flux to refer to the corrected MAXI photon flux under the assumption of a Crab-like spectrum.  

The dataset includes flare-like profiles due to unexpected background increases and dip-like profiles caused by the shadow effects of solar panels\footnote{MAXI Readme: \url{http://maxi.riken.jp/top/readme.html}}. Calibration methods also introduce variability in 2--4~keV keV flux measurements due to absorption effects, which can impact the accuracy of flux values in this energy range. Additionally, uneven coverage arises from factors such as the source's proximity to the Sun and system maintenance downtime, affecting data availability for specific binary orbits.  

\section{Flux and hardness average trends}
\label{section:overall_data_distribution_analysis} 

In this section, we first introduce some relevant definitions for the entirety of our analysis. Subsequently, we explain the determination of our average trends and describe their evolution throughout the orbital period. These curves not only provide valuable information about the large-scale structures, as we discuss in Sect.~\ref{subsection:large_scale_steady_structures}, but also serve as a reference for quantifying orbit-to-orbit variability (see Sect.~\ref{section:orbital_profiles_analysis}).  

\subsection{Preliminary definitions}
\label{subsection:preliminary_definitions}

In the subsequent analysis, we designate the 2--4~keV energy range as the ``soft band'' and the 4--10~keV as the ``hard band''. The impact of absorption on the expected fluxes in these adjacent bands is markedly different for the range in $N_\textrm{H}$ typical for Vela X-1, i.e. 1--100$\times$10$^{22}\mathrm{cm}^{-2}$ \citep[][Fig.~4a]{Kretschmar_2021a}, where it mainly impacts low energies, as shown in Fig.~\ref{figure:xmm}.  

The orbital phases ($\phi_\mathrm{orb}$) are obtained with the ephemeris from \citet{Kreykenbohm_2008}. We employ the mid-eclipse time of 52974.227$\pm$0.007~MJD as the phase zero, and 8.964357$\pm$0.000029 day as the orbital period.  

The hardness ratio of the selected soft and hard energy bands is used to quantify absorption variability effects. We define the hardness ratio as: 
\begin{equation}\mathrm{hardness\,ratio} = (\mathrm{H} -\mathrm{S})/(\mathrm{H}+\mathrm{S}),\end{equation}
where H, and S represent the MAXI fluxes in hard and soft band, respectively. 

Unlike the traditional eclipse definition spanning from orbital phases 0.89 to 0.1 \citep[e.g.,][]{Kreykenbohm_2008}, our study finds a notable decrease in hard flux from phases 0.87 to 0.89, making it difficult to establish a hardness ratio. Previous studies have shown that the definition of orbital eclipse phases is energy-dependent \citep[e.g., Fig.~3 in][]{Falanga:2015}. As a consequence, we restrict our analysis to the orbital phase range [0.10, 0.86], which encompasses 53.9$\%$ of the total sample of soft and hard fluxes.
   
\begin{figure*} 
\centering
\begin{subfigure}{0.49\textwidth} 
\centering
\includegraphics[width=1.0\textwidth]{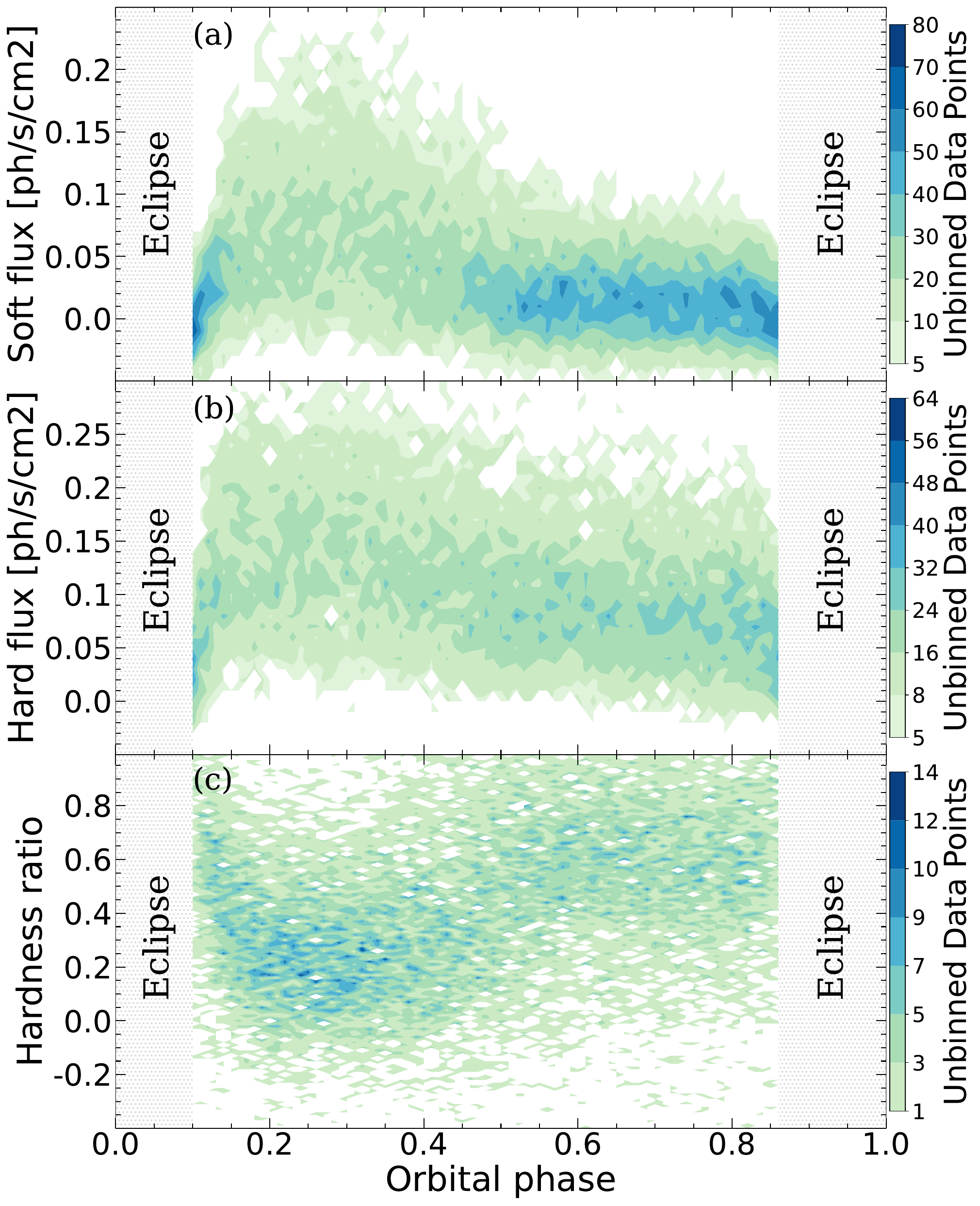}
\end{subfigure}
\hfill 
\begin{subfigure}{0.455\textwidth} 
\centering 
\includegraphics[width=1.0\textwidth]{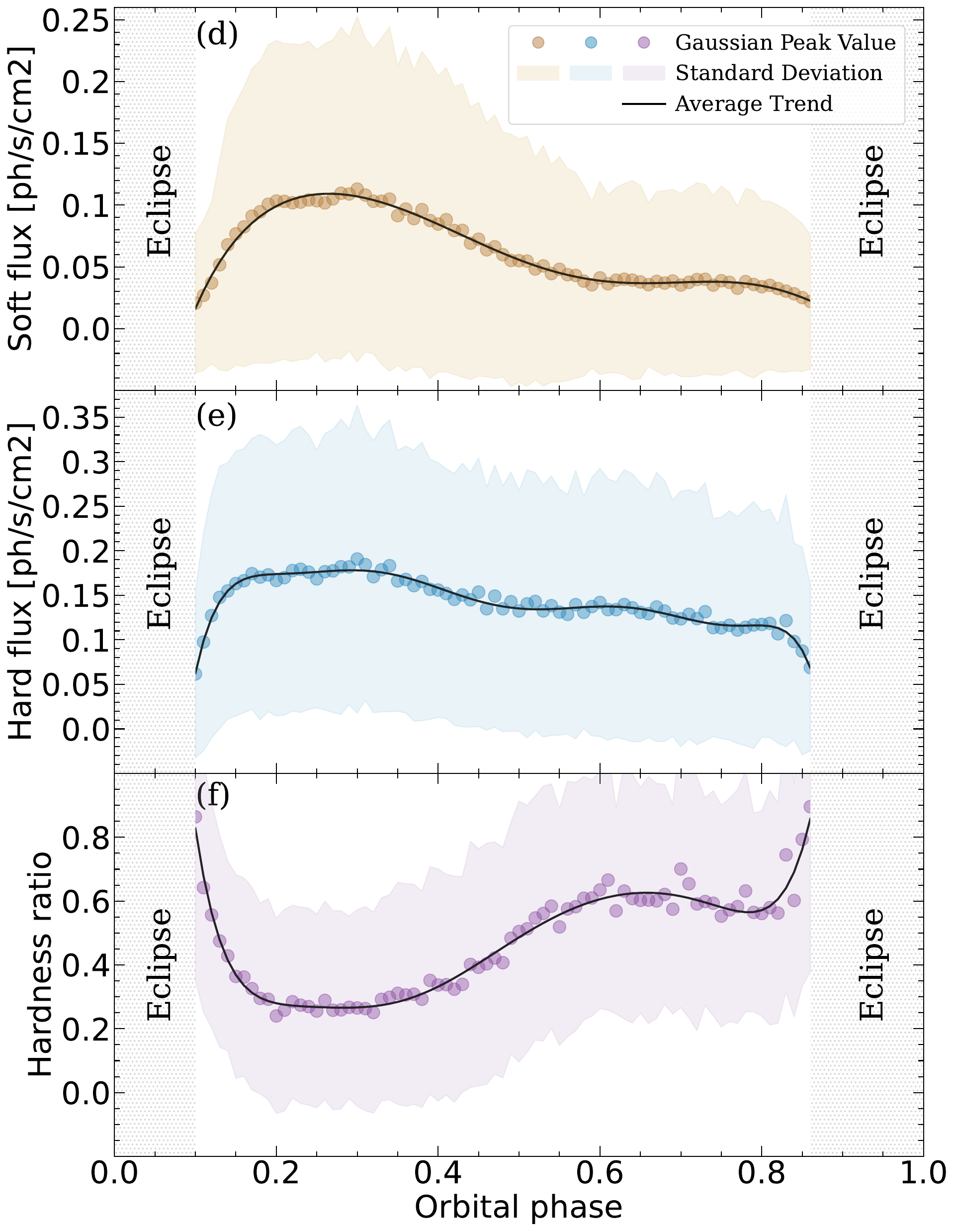}
\end{subfigure}
\caption{
\textit{Left}: 2D histograms showing the unbinned MAXI/GSC dataset, depicting the flux in the 2--4 keV (soft band; top), and 4--10 keV energy range (hard band; middle), and the derived hardness ratio (bottom), calculated as explained in Sect.~\ref{section:overall_data_distribution_analysis}. The number of unbinned data points is colour-coded, where lighter shades denote smaller values and darker shades indicate higher ones. \\ \textit{Right}: average trends and their associated standard deviations derived from fits to the 2D histograms. Each shaded region is specifically colour-coded for visual association with Figs.~\ref{figure:orbits} and~\ref{figure:lightcurve}. 
For further details, we refer to the text. \\ 
}
\label{figure:trends}
\end{figure*}

\begin{figure} 
\centering
\includegraphics[width=0.49\textwidth]{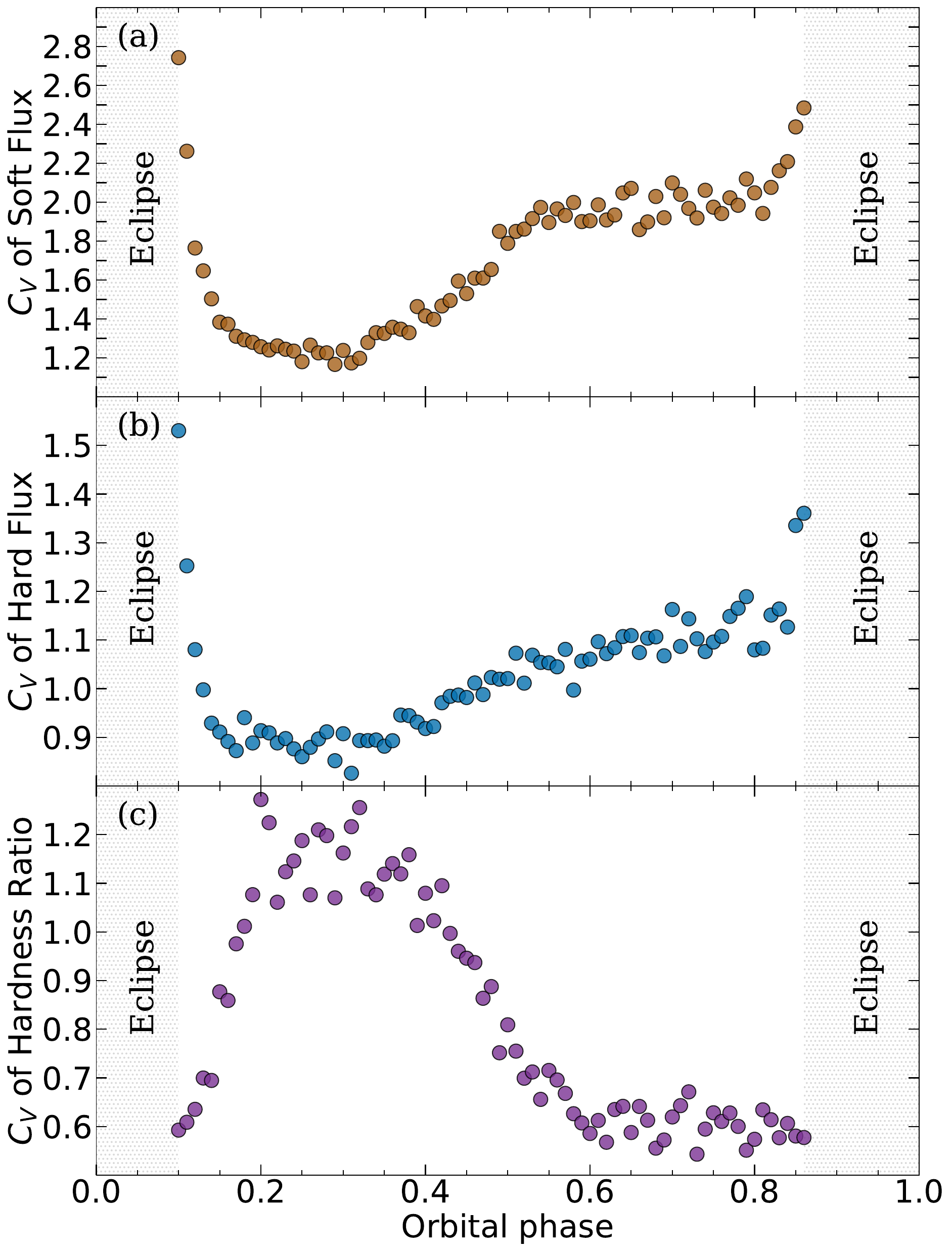}
\caption{
Coefficient of variation ($C_\mathrm{V}$, see Eq.~\ref{equation:CV}) of the flux and hardness average trends from Fig.~\ref{figure:trends} (right panels). 
}
\label{figure:cv}
\end{figure}

\subsection{Average trend analysis}
\label{subsection:flux_and_hardness_ratio_average_trends}

The average trends in Fig.~\ref{figure:trends} (right) are derived from the histograms shown in the left panels of the same figure. Each orbital phase bin is of a size of 0.01, and is fitted using a normalised Gaussian distribution and a constant term. Due to sparse counts in each phase bin, the use of Cash statistics \citep{Cash_1979} ensures non-biased fitting. The 1$\sigma$ uncertainties in the Gaussian parameters are derived through Markov Chain Monte Carlo sampling. The peak of the Gaussian represents the mean hardness of the phase bin, and its standard deviation provides the 1$\sigma$ uncertainty. The fitting results for all phase bins are displayed as a function of the orbital phase. Moreover, an eighth-degree polynomial is applied to capture the overarching pattern, providing an overview of the dataset's average behaviour.  

To better understand the relative variability in each average trend curve across orbital phases, we compute the coefficient of variation ($C_\mathrm{V}$) by normalising each standard deviation ($\sigma$) to the associated mean value ($\mu$) as follows:

\begin{equation} \label{equation:CV}
C_\mathrm{V} = \frac{\sigma}{\mu}.
\end{equation}

This metric enables direct comparisons across flux and hardness ratio curves, revealing variability patterns and trends relative to changes in their mean. In Fig.~\ref{figure:cv}, we display the $C_\mathrm{V}$ for the fluxes and hardness ratio curves from Fig.~\ref{figure:trends} (right panels). We refer to Sect.~\ref{subsection:description_flux_and_hardness_average_evolution} for further details in the $C_\mathrm{V}$ behaviour.

\subsection{Description of the flux and hardness average evolution} 
\label{subsection:description_flux_and_hardness_average_evolution}

We observe a consistent decline with the orbital phase in both the soft and hard flux average trends (Fig.~\ref{figure:trends} d,e), with a more significant decrease observed in the soft energy band. This is attributed to its heightened sensitivity to absorption effects, evidenced by an $\sim$90$\%$ decrease between the first and second halves of the orbital phase. The hard band is limited to the 4--10 keV energy range and is still noticeably affected by absorption (Fig.~\ref{figure:xmm}). The hardness average trend (Fig.~\ref{figure:trends}f) exhibits lower hardness values in the first half compared to the second one. 

While the flux mean decreases with the orbital phase, the associated standard deviation decreases less, causing the coefficient of variation (Eq.~\ref{equation:CV}) to increase and eventually plateau after orbital phase 0.5 (Fig.~\ref{figure:cv}a,b), indicating greater variability in the first half of the orbit. This is illustrated by the bump in the coefficient of variation of the hardness mean profile at orbital phases 0.2–0.4 (Fig.~\ref{figure:cv}c). Further discussion on this topic can be found in Sect.~\ref{subsubsection:stochastic_variability_from_the_average_behavior}. 

\section{Orbit-to-orbit variability in hardness profiles}
\label{section:orbital_profiles_analysis} 

In the following section, we analyze the variability of hardness ratios at the level of individual binary orbits, referred to hereafter as hardness profiles. First, we explain the construction of our sample of hardness profiles (Sect.~\ref{subsection:binary_orbit_selection}). Thereafter, we describe the statistical method developed for the classification and quantification of the observed variability (Sect.~\ref{subsection:hardness_profile_classification}).  

\subsection{Individual hardness profile sample selection}
\label{subsection:binary_orbit_selection} 

Our analysis begins by extracting the flux outside the eclipse region to derive soft and hard fluxes binned per day, after which we compute hardness ratios. In Fig.~\ref{figure:lightcurve}, we showcase an example of MAXI/GSC light curves, where median values are calculated within each time interval.  

\begin{figure} 
\centering
\includegraphics[width=0.49\textwidth]{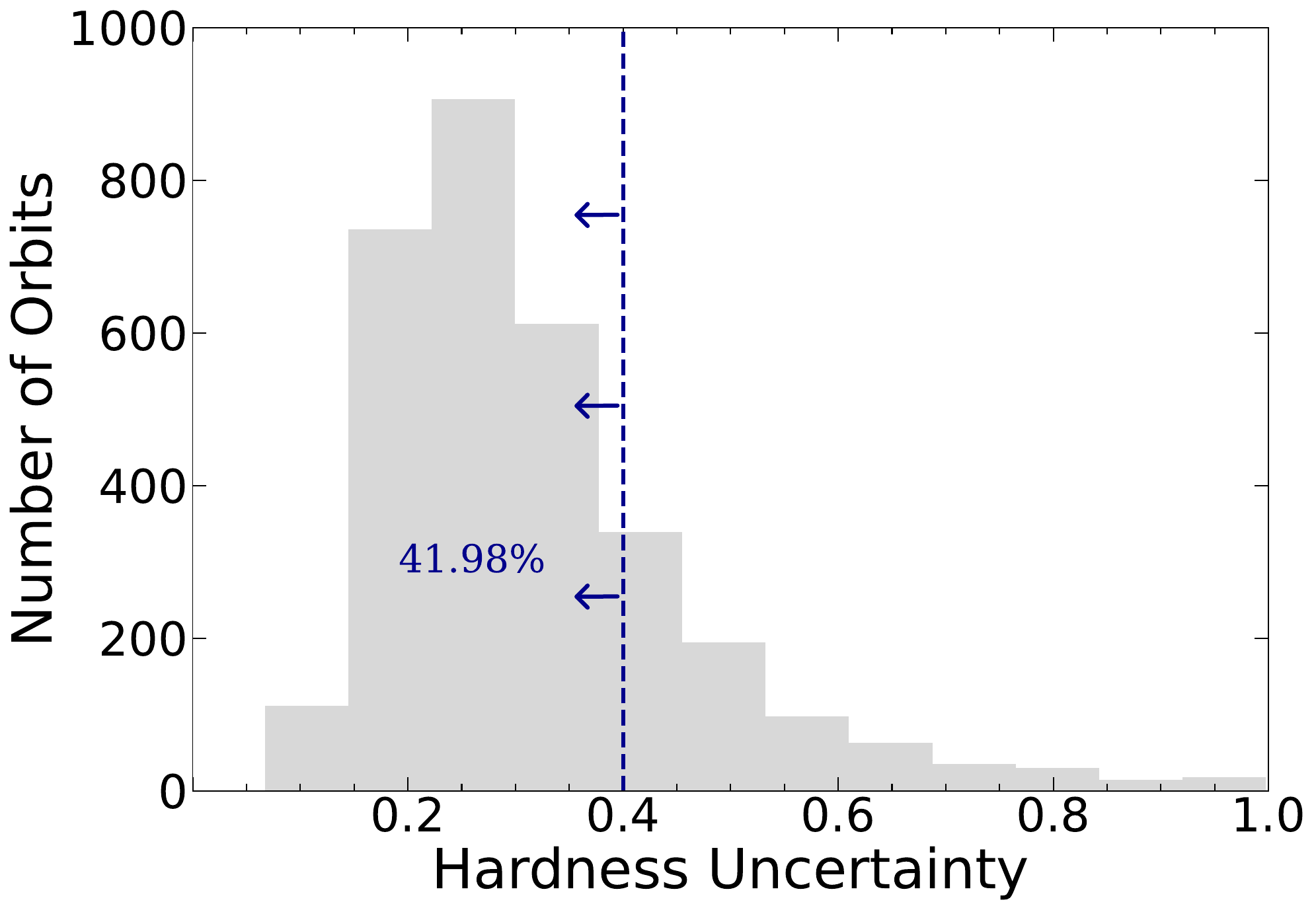}
\caption{
Histogram of hardness uncertainties derived from fluxes outside the eclipse region, as defined in Sect.~\ref{subsection:preliminary_definitions}. The threshold, indicated by a vertical line refines the sample by excluding hardness values with uncertainties exceeding 0.4. See further details in Sect.~\ref{subsection:binary_orbit_selection}.
}
\label{figure:errHR_histogram}
\end{figure}

The investigation into orbit-to-orbit variability relies on hardness profiles with low uncertainties. Therefore, data points with uncertainties exceeding 0.4 are excluded. This criterion is informed by the uncertainty histogram depicted in Fig.~\ref{figure:errHR_histogram}. The histogram shows that imposing a stricter filter significantly diminishes the sample size without commensurately enhancing the accuracy of hardness ratios. This refinement process leads to a 42.0$\%$ reduction in the initial sample size.  

Lastly, to ensure sufficient information for assessing hardness variability in each binary orbit, we prioritise achieving adequate data coverage. Specifically, we refer to the system’s orbital period (Sect.~\ref{subsection:preliminary_definitions}) to guarantee a minimum of five 1-day-binned hardness values per orbit, capturing more than half of the maximum available information within each binary orbit. As a result, the final hardness profile sample size comprises 30.5$\%$ of the initial sample, totalling 315 orbital hardness profiles.  

\subsection{Deciphering orbit-to-orbit variability}
\label{subsection:hardness_profile_classification}

The quantification of orbit-to-orbit hardness variability is not a trivial task due to the complex nature of the system. We quantify the relative variability of every hardness ratio value within every hardness profile (Sect.~\ref{subsection:quantifying_variability_with_CDF}). Following this, we propose a criterion for classifying the evolution of hardness ratios in binary orbits (Sect.~\ref{subsection:streamlined_classification}). Our statistical approach facilitates the understanding of individual hardness profiles, despite the resolution limitations of our dataset. 

\subsubsection{Quantifying relative variability using the complementary Cumulative Distribution Function}
\label{subsection:quantifying_variability_with_CDF}

The first step in the classification process involves evaluating every hardness ratio within each hardness profile, along with its corresponding uncertainty, relative to the hardness average trend. To achieve this, we utilise the Cumulative Distribution Function (CDF), which describes the probability that a random variable takes on a value less than or equal to a specific value. In our context, it helps determine how the observed hardness ratios compare to the average trend by considering their uncertainties.

Treating the hardness ratio as the mean and its associated uncertainty as the 1$\sigma$ standard deviation, we calculate the fraction of the Gaussian distribution (which represents the uncertainty in the hardness ratio) that lies above the hardness average trend by using the complementary CDF, i.e., (1$-$CDF).

For example, hardness values with (1$-$CDF) values of 0.5 indicate that half of the Gaussian distribution is above and half is below the average trend. Conversely, (1$-$CDF) values close to 1 indicate that most of the distribution is above the average trend, while those nearing 0 indicate that most of the distribution is below the average trend. 

\subsubsection{Streamlined classification: median of complementary CDF values for each hardness profile}
\label{subsection:streamlined_classification}

\begin{figure} 
\centering
\includegraphics[width=0.49\textwidth]{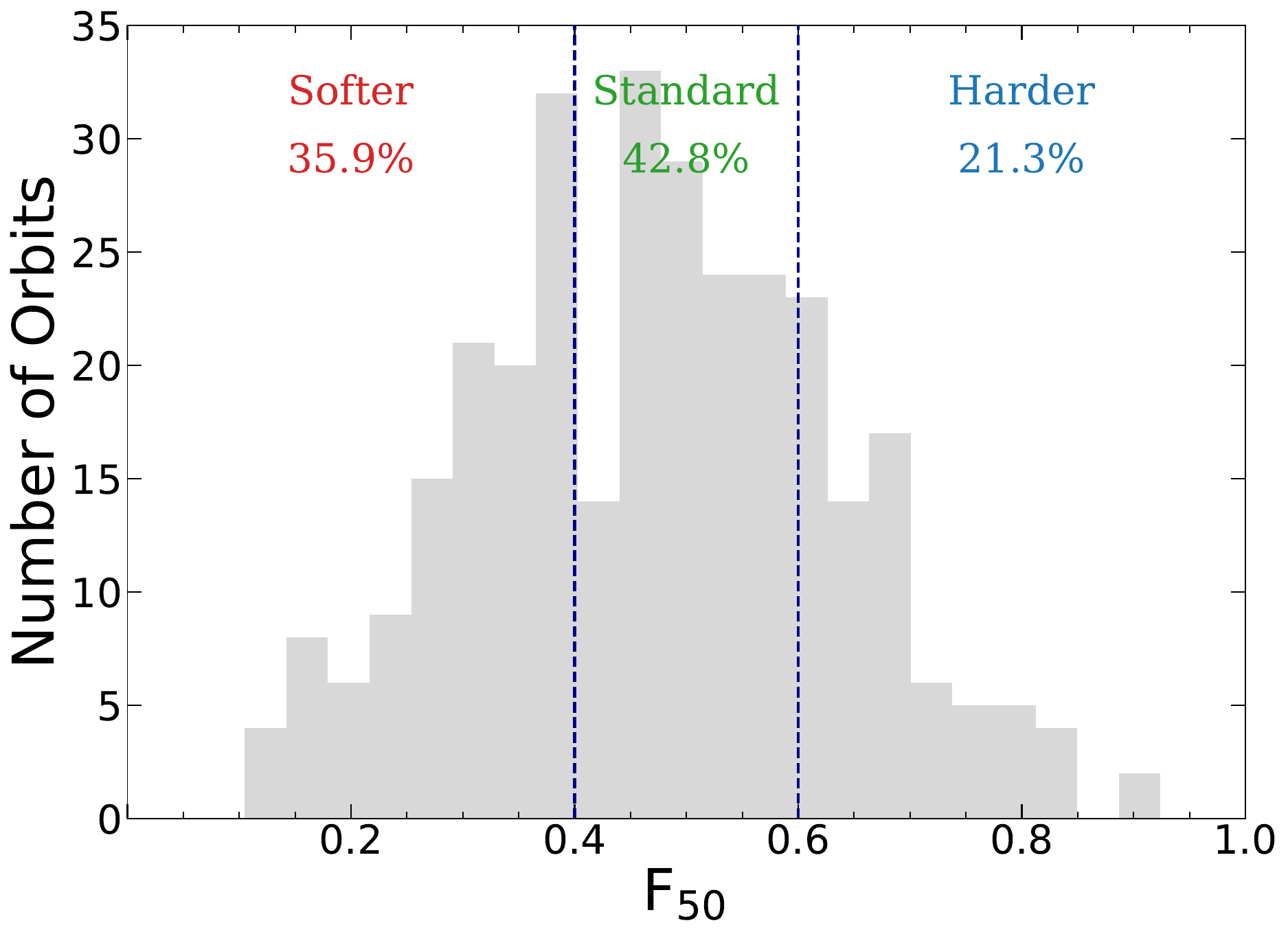}
\caption{
Histogram of the $F_\mathrm{50}$ parameter (Eq.~\ref{equation:F_50}) illustrating the streamlined classification as explained in Sect.~\ref{subsection:streamlined_classification}, with vertical dashed blue lines indicating the defined thresholds for each type of hardness profile. We refer to the text for further details.
}
\label{figure:histogram_median}
\end{figure}

After computing the (1$-$CDF) for each hardness ratio (Sect~\ref{subsection:quantifying_variability_with_CDF}), we calculate the median of these values within each hardness profile to establish the parameter $F_\mathrm{50}$, as follows:
\begin{equation} \label{equation:F_50}
F_\mathrm{50} = \mathrm{median}\sum_{i=1}^{N} (1-CDF)_i,
\end{equation}
where $N$ is the number of hardness ratios in the binary orbit, and (1$-$CDF) is the complementary CDF associated with each hardness ratio. We associate an $F_\mathrm{50}$ parameter with each binary orbit in our sample. 

Then, we classify each hardness profile based on its $F_\mathrm{50}$ according to the following criteria: orbits with an $F_\mathrm{50}$ falling within $0.4 \leq F_\mathrm{50} \leq 0.6$ are classified as ``standard'', while those with $F_\mathrm{50} > 0.6$ are defined as ``harder'' profiles, and those with $F_\mathrm{50} < 0.4$ are categorised as ``softer'' profiles. The $F_\mathrm{50}$ histogram of the analysed profiles is displayed in Fig.~\ref{figure:histogram_median}, depicting the defined thresholds of the criteria. This classification reveals that 42.8$\%$ of the orbits align with the mean profile and are classified as standard profiles, while 35.9$\%$ are softer, and 21.3$\%$ are harder profiles. The analysis of this distribution is further discussed in Sect.~\ref{subsection:absorption_variability_in_orbit_to_orbit_hardness_profiles}.

\begin{figure*} 
\centering
\begin{subfigure}{0.47\textwidth}
\centering
\includegraphics[width=1.0\textwidth]{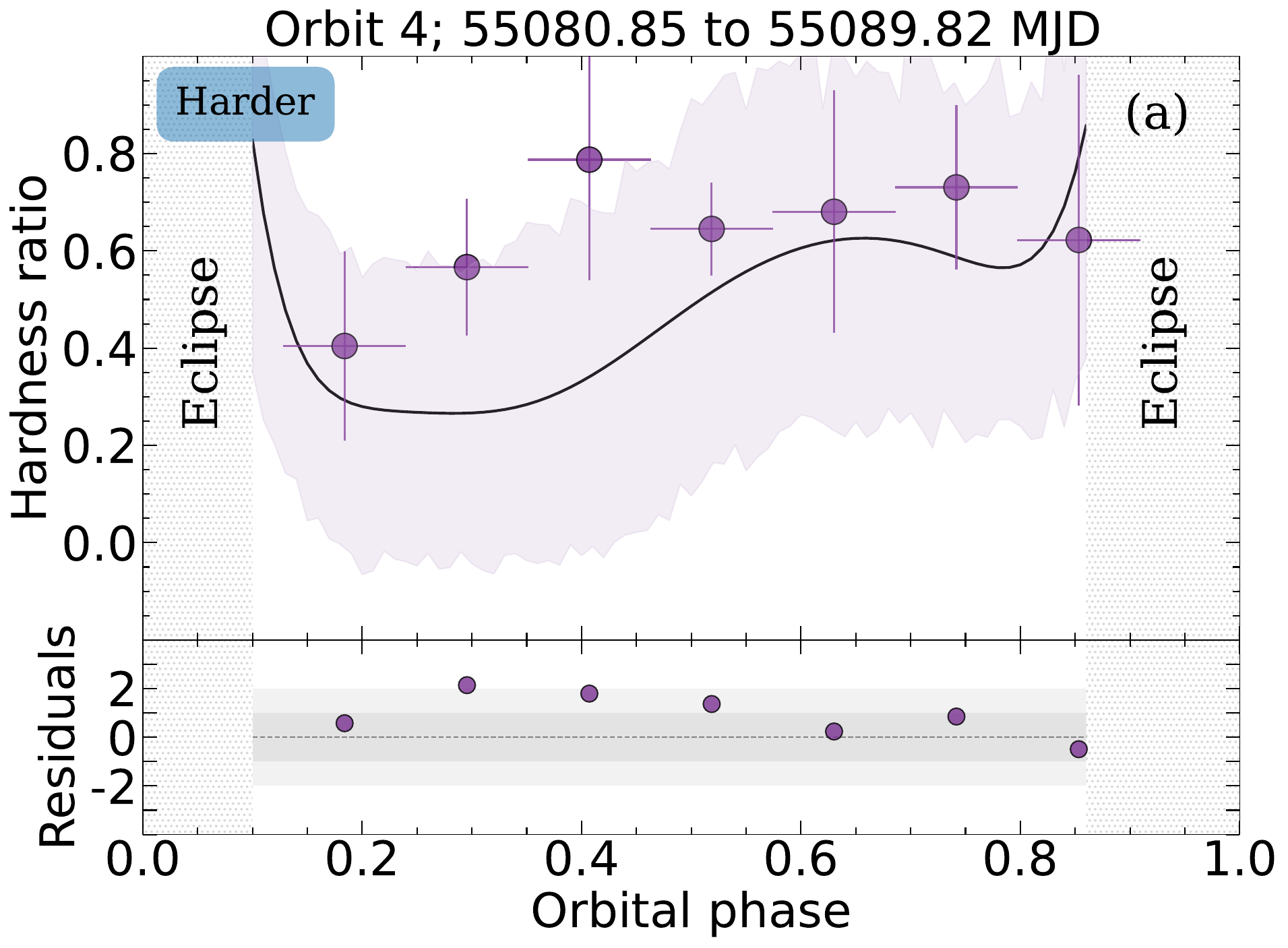} \label{figure:trends-a}
\end{subfigure}
\hfill
\begin{subfigure}{0.47\textwidth} 
\centering
\includegraphics[width=1.0\textwidth]{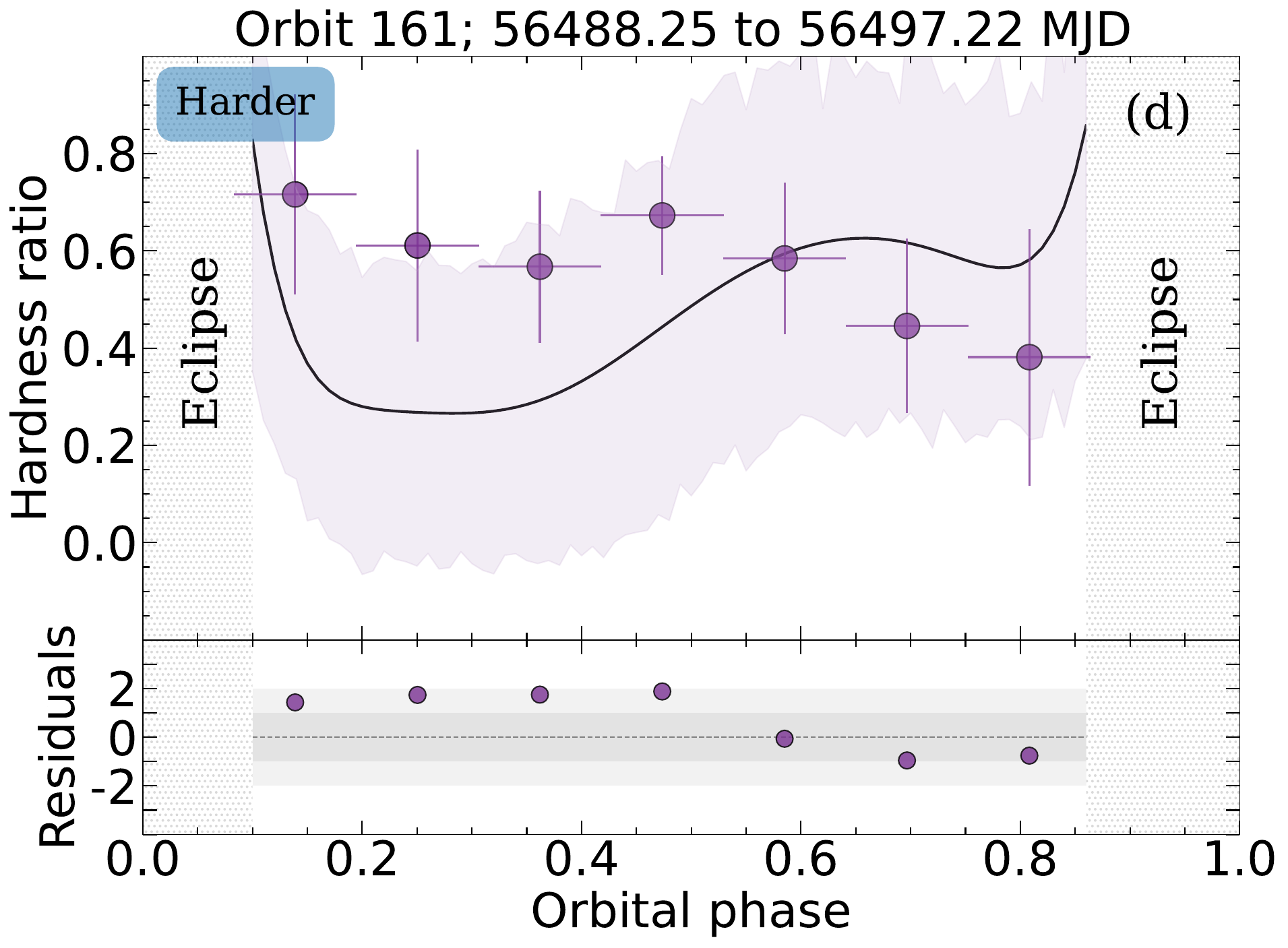} \label{figure:trends-d}
\end{subfigure}
\hfill
\begin{subfigure}{0.47\textwidth} 
\centering
\includegraphics[width=1.0\textwidth]{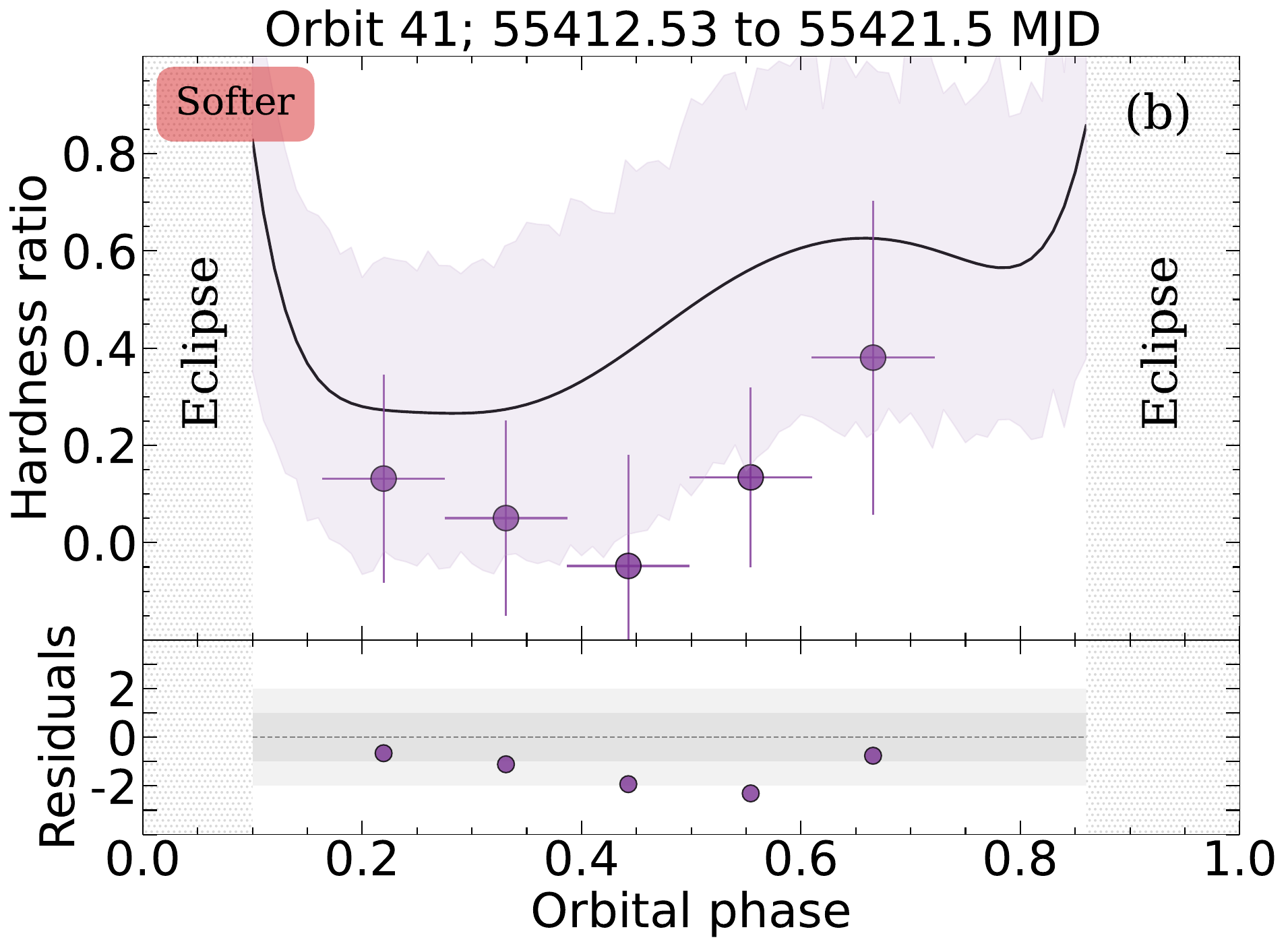} \label{figure:trends-b}
\end{subfigure}
\hfill
\begin{subfigure}{0.47\textwidth} 
\centering
\includegraphics[width=1.0\textwidth]{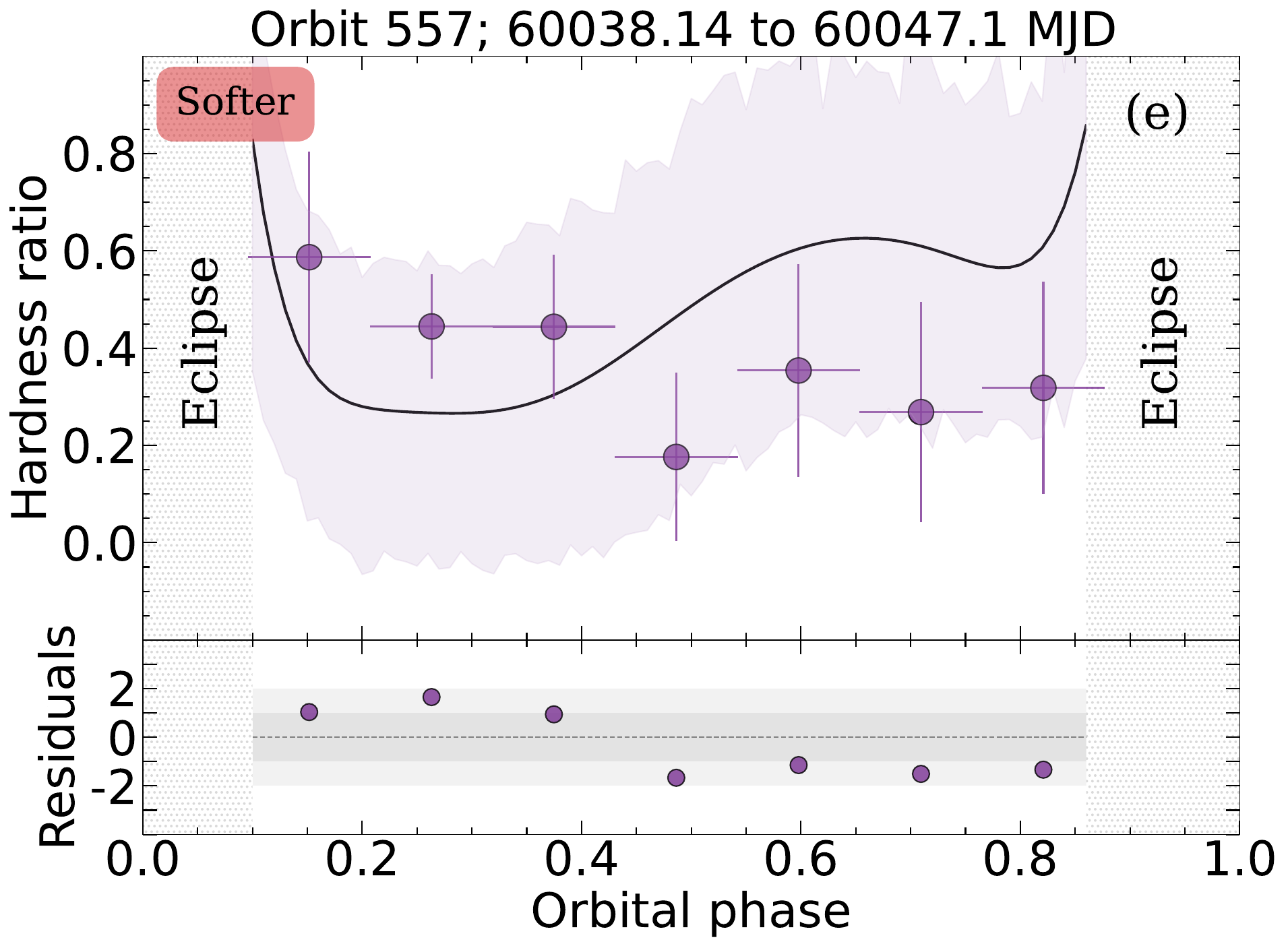} \label{figure:trends-e}
\end{subfigure}
\hfill
\begin{subfigure}{0.47\textwidth} 
\centering
\includegraphics[width=1.0\textwidth]{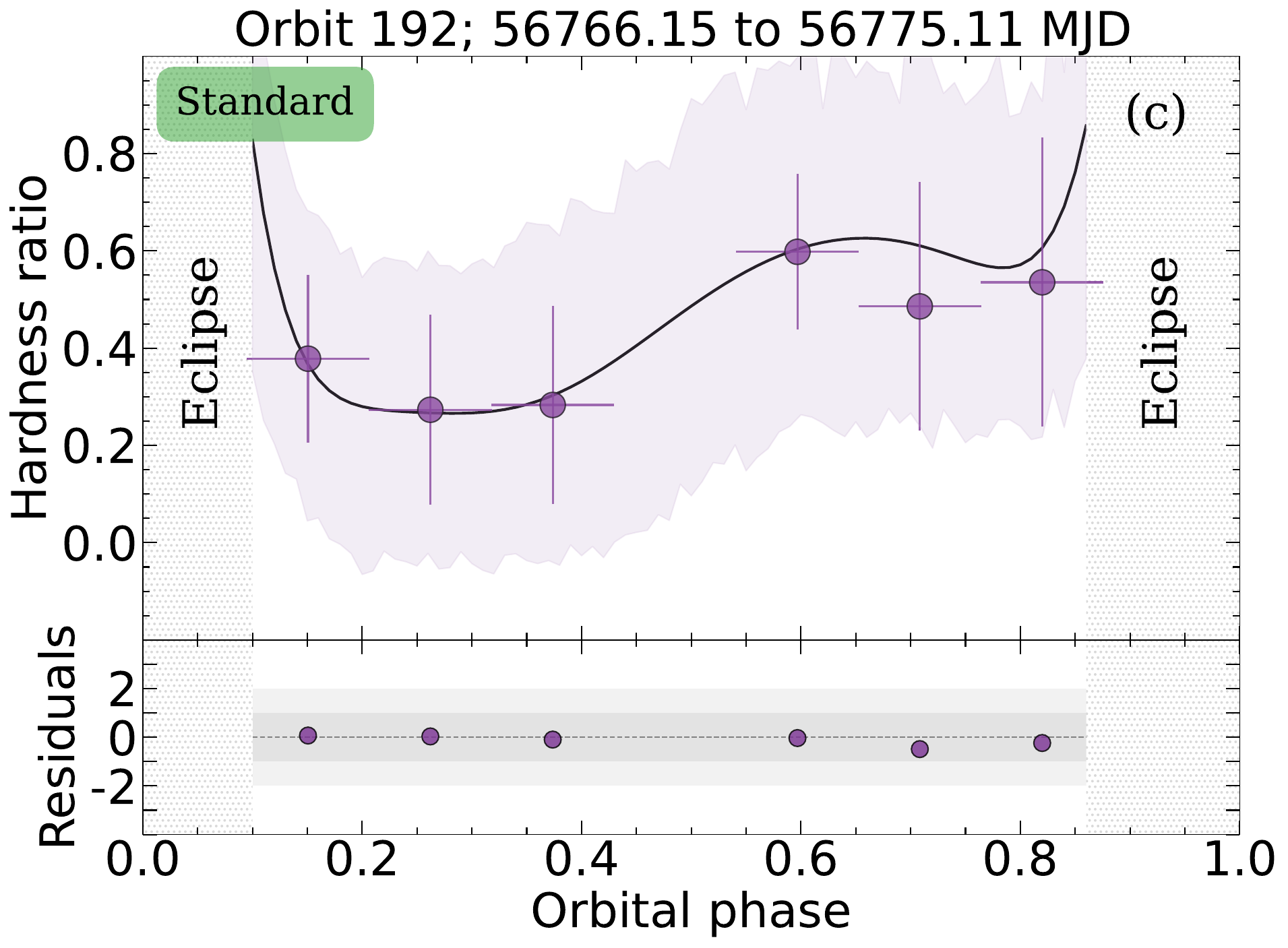} \label{figure:trends-c}
\end{subfigure}
\hfill
\begin{subfigure}{0.47\textwidth} 
\centering
\includegraphics[width=1.0\textwidth]{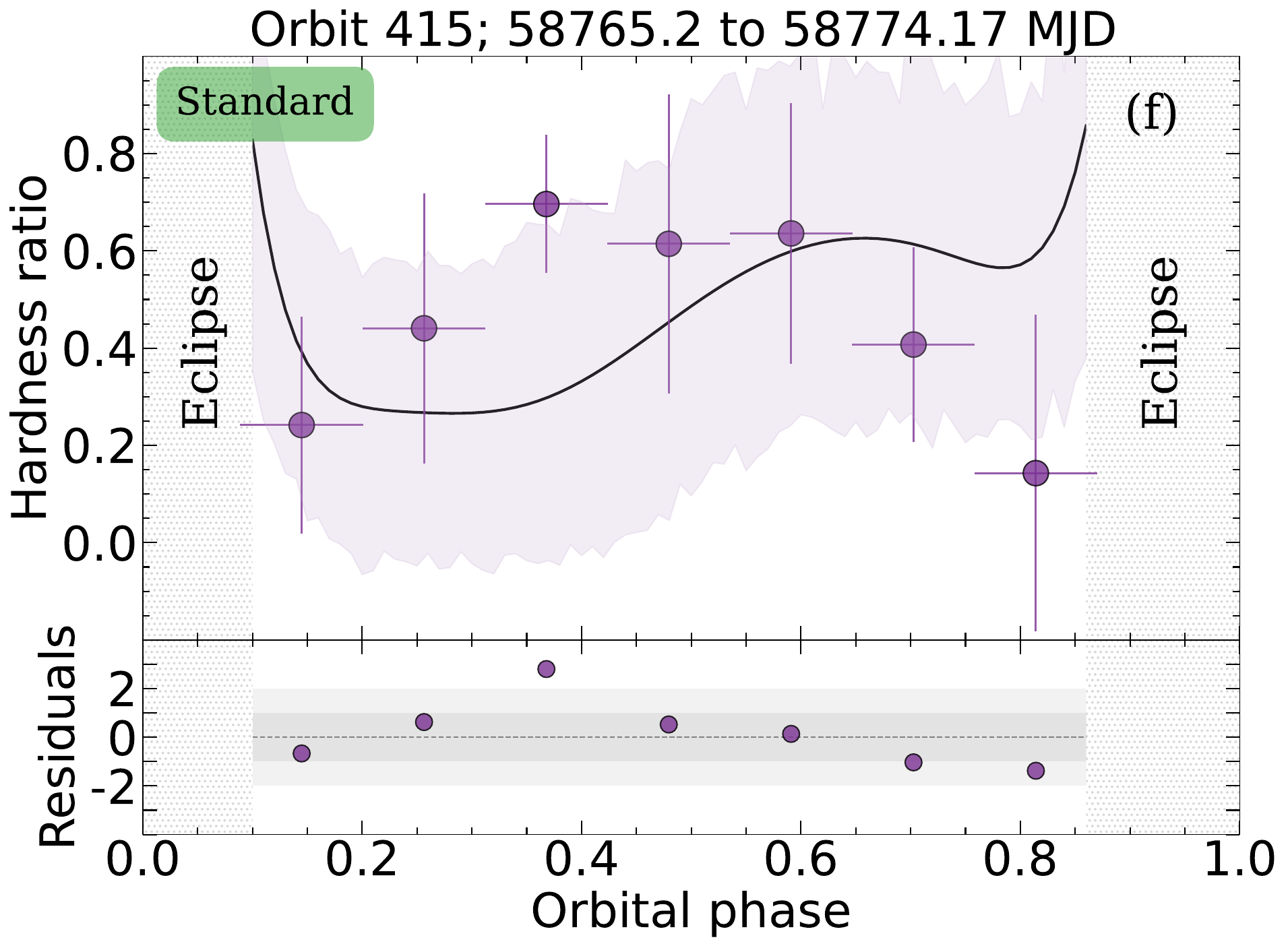} \label{figure:trends-f}
\end{subfigure}\caption{
Orbit-to-orbit hardness profiles categorized into three groups based on F$_{\mathrm{50}}$ (Eq.~\ref{equation:F_50}) thresholds: harder ($F_{\mathrm{50}}>0.6$, top, blue), softer ($F_{\mathrm{50}}<0.4$, middle, red), and standard ($0.4\geq\mathrm{F}_{50}\geq0.6$, bottom, green). 
The left column showcases well-defined orbits, while the right column displays orbits from the same group with larger variability (see Appendix~\ref{appendix:comprehensive_variability_quantification} for a more elaborated statistical analysis derived from Sect.~\ref{subsection:streamlined_classification}). We refer to the text for further details. 
}
\label{figure:orbits}
\end{figure*}

In Fig.~\ref{figure:orbits}, we show two representative hardness profile examples from each of the three categories. The left column displays hardness evolution patterns, which one would unequivocally classify visually in the same manner as our statistical criterion. However, the simplicity of our classification method is demonstrated by the hardness evolution patterns in the right column, where visual classification is less evident. 

In Appendix~\ref{appendix:comprehensive_variability_quantification}, we elaborate on our approach to quantifying variability using the three-group classification based on the $F_\mathrm{50}$ parameter (Eq.~\ref{equation:F_50}). This method measures the magnitude and direction of outliers from the overall evolution for each hardness profile, providing a visual and comprehensive description of the hardness ratio evolution in each binary orbit without the need for individual sample inspection (Fig.~\ref{figure:binary_orbit_map}).

\section{Discussion}
\label{section:discussion} 

In Sect.~\ref{section:overall_data_distribution_analysis}, show the Vela X-1 long-term evolution, revealing persistent patterns over more than 14 years of continuous MAXI monitoring (as also found by several previous studies). However, significant variability in X-ray absorption is still evident (Fig.~\ref{figure:trends}). The fluctuations in the hardness ratio are most pronounced during periods of low absorption between orbital phases 0.2--0.4 (Fig.~\ref{figure:cv}), but also occasionally high $N_{\rm H}$ (Fig.~\ref{figure:nh}). This section discusses explanations for the absorption variability for individual binary orbits. Additionally, we discuss the connection between the spin state and orbit-to-orbit absorption variability.  

\subsection{Hardness as a diagnostic to trace absorption variability} 
\label{subsection:hardness_traces_absorption}
To justify the use of hardness ratios and demonstrate their sensitivity to variations in $N_{\rm H}$, we simulate spectra based on the best-fit model described in \citet{Diez_2023}, covering the range of spectral slopes observed in their study. Assuming minimal absorption, when the continuum slope has the most significant effect, the hardness ratio difference between photon index extremes (0.8 and 1.2) is only $\sim$0.13. Conversely, assuming the hardest continuum, where $N_{\rm H}$ has the least impact on the hardness ratio, and varying $N_{\rm H}$, the difference exceeds 0.5. For all other cases the slope's impact is even less significant, indicating that absorption variations primarily dominate the observed hardness ratios.  

Variations in the amount of absorbing material due to the non-uniform stellar wind or the accretion wake structure cause significant energy-dependent flux changes in the energy bands covered by MAXI (Sect.~\ref{section:introduction} and  Fig.~\ref{figure:xmm}), leading to noticeable hardness ratio variations and allowing us to use hardness ratio variations as a proxy for absorption changes.

In the analysis, the changing ionisation structure of the wind, particularly during flares, is addressed by considering the variability in the ionisation parameter, $\xi$, which is influenced by X-ray luminosity (L$_{\mathrm{X}}$), distance from the neutron star, and local number density. Simulations assume a constant L$_{\mathrm{X}}$, resulting in $\xi$ primarily depending on distance and density. Flares, which last for thousands of seconds and can increase L$_{\mathrm{X}}$ by a factor of 3-4 and more, can disrupt this assumption as the ionisation parameter and the local density can fluctuate substantially. However, the time resolution of MAXI, which provides a $\sim$90-second snapshot every $\sim$90 minutes, limits the ability to capture these rapid changes. As flares are often longer than the MAXI sampling interval, they introduce scatter in the average measurements rather than detailed temporal profiles. This means that the variations in ionisation due to flares are averaged out in the analysis, contributing to the observed scatter in the hardness ratio.

Our study focuses on changes within individual binary orbits. In Fig.~\ref{figure:nh}, we compare the long-term average hardness ratio trend from our MAXI/GSC sample (Fig.~\ref{figure:trends}f) with $N_{\rm H}$ values obtained by various X-ray observatories. 
The average hardness trend has been arbitrarily scaled to facilitate comparison and emphasise similarity in overall trend. Direct comparisons of absolute $N_{\rm H}$ values from various studies should be approached cautiously, given distinct models and assumptions in the literature. Nonetheless, Fig.~\ref{figure:nh} illustrates the variability in $N_{\rm H}$ measurements obtained through pointed X-ray observations at comparable orbital phases over different time periods. 
The close correspondence between our average hardness profile as a function of orbital phase and the literature measurements shows that monitoring over multiple orbits is necessary to properly track the evolution of the absorption components. 

\subsection{Large scale steady structures}
\label{subsection:large_scale_steady_structures} 

The mean hardness profile (Fig.~\ref{figure:trends}f) reveals an asymmetry between the first and second halves of the orbital period, suggesting the presence of large-scale structures, such as an accretion wake trailing the neutron star. The wake structure is attributed to the gravitational and radiative impacts of the neutron star on the outflowing stellar wind \citep[see][]{Kretschmar_2021a}. The soft flux average trend reveals a significant reduction of $\sim$90$\%$  between the first and second halves of the orbit, coinciding with the presence of the wake structure in the line of sight. A similar trend is observed in the average profile of the hard flux (Fig.~\ref{figure:trends}e), albeit less pronounced since the hard flux band is less affected by absorption. Early studies by \citet{Bessell_1975} and subsequent simulations by \citet{Blondin_1991} and \citet{Manousakis_2012} highlight the persistent trailing of this wake throughout the neutron star's orbit. 

During the eclipse egress, occurring at orbital phases 0.1--0.2, there is a rapid decrease in the hardness ratio, also indicated by the $N_\mathrm{H}$ values in Fig.~\ref{figure:nh}. This decline is typically linked to the presence of the extended stellar atmosphere obscuring the line of sight \citep{Sato_1986a, Lewis_1992a}. Next, a plateau emerges at orbital phases 0.2--0.4. This likely arises from the wake structure positioned away from our line of sight, so that our line of sight intersects a wind region of lower density leading to lower observed hardness values. 

Thereafter, the average hardness smoothly increases between orbital phases 0.4 to 0.6. Prior observational results and theoretical models show steep increases for individual orbits: \citet{Ohashi_1984a}, using Tenma data, finds that the $N_\mathrm{H}$ steeply increases within this phase interval (Fig.~\ref{figure:nh}). \citet{Diez_2023} analyse an \xmmnewton observation spanning orbital phases 0.34--0.48 (Fig.~\ref{figure:nh}) and show a steep increase, but at a slightly different point in the orbit. Similarly, \citet{Manousakis_2012} predict a model with a similar steep absorption increase, although it was developed for a different system with similar system parameters. The smooth continuous trend observed with MAXI during these orbital phases can then be attributed to variations in the onset timing of absorption increase across different orbits, resulting in an overall smoother average transition. 

A second plateau with sustained higher hardness values is observed from 0.6 to 0.8, attributed to our line of sight intersecting the wake. However, during this second plateau, the hardness average trend shows a slight decrease due to the interplay of two phenomena: the trailing wake structure begins to move out of our line of sight, resulting in less material absorbing the X-ray emission of the neutron star and simultaneously, the extended stellar atmosphere of the companion starts absorbing the X-ray emission. The decrease caused by the displacement of the wake structure is partially counteracted by the onset of eclipse ingress. However, it is only after phase 0.8 that the absorption levels of the eclipse ingress are sufficiently high to produce a similar rapid absorption increase, mirroring the eclipse egress behaviour. 

\subsection{Stochastic variations in absorption}
\label{subsection:stochastic_variability_from_the_average_behavior}

The dynamic interaction among the different components of the Vela X-1 system, including the neutron star, the stellar wind, and the accretion wake results in stochastic fluctuations observed in X-rays. First, we examine the origins of this stochastic variability based on literature (Sect.~\ref{subsection:context_from_previous_work}). We then deduce it through the analysis of MAXI long-term average trends (Sect.~\ref{subsubsection:stochastic_variability_from_the_average_behavior}), as well as from individual hardness profiles (Sect.~\ref{subsection:absorption_variability_in_orbit_to_orbit_hardness_profiles}).

\subsubsection{Previous observational results and model descriptions}
\label{subsection:context_from_previous_work}

Multiple components can contribute to the  stochastic variability, such as variations in material integration along the line of sight induced by clumps in the stellar wind, accumulation of matter and modifications of the flow geometry at the outer rim of the neutron star magnetosphere, or changes in opacity due to changes in wind ionisation structure.  

Wind clumps intercepting the line of sight can produce absorption variability on short timescales. The 2D model from \citet{Oskinova_2012} considers massive and intermediate-sized overdense shells propagating radially from the donor star. They compute the time-dependent extinction coefficient, revealing that radially compressed clumps exhibit greater variability than their spherical counterparts. \citet{El_Mellah_2020a} conduct a comprehensive exploration of column density variability induced by a clumpy wind assuming spherical clumps. They develop a 3D model of radial clump flow computing median $N_{\rm H}$ orbital profiles and standard deviations over multiple orbital periods. The $N_{\rm H}$ variation timescale was found to be approximately given by the flyby time of the smallest clumps along the line of sight, providing insights into their spatial extent. Moreover, the authors show that the clump mass can be constrained using the column density standard deviation. 

Changes in the flow structure near the neutron star's magnetosphere contribute to the formation of wakes, which in turn affect the variability of absorption along the line of sight. However, the size of the wake structure exceeds not only the orbital separation between the neutron star and its donor, but also significantly surpasses the size of the neutron star's magnetosphere \citep[][Fig.~2]{Kretschmar_2021a}. Consequently, while flow structure changes do play a role in stochastic absorption variability, their impact is minor. The smooth stellar wind dynamics in hydrodynamics numerical simulations by \citet{Manousakis_Walter_2011} successfully replicate the column density excesses observed in the orbital phases 0.4--0.8, attributing them to an overdense accretion wake trailing the neutron star. \citet{Malacaria_2016a} demonstrate the inhomogeneous nature of this wake structure through detailed spectral analysis of enhanced absorption events near inferior conjunction, in agreement with \citet{Doroshenko_2013}.  

In the vicinity of the neutron star, extending from the magnetosphere to regions approximately 100 times farther, intrinsic variations in X-ray ionising emissions influence the structure of the stellar wind \citep[][Fig~2]{Kretschmar_2021a}. In highly ionised regions, particularly those near the neutron star or in low-density areas, X-rays can pass through with minimal absorption. Consequently, fluctuations in the wind's ionisation state can lead to observed variations in the column density, even if the actual amount of absorbing material along the line of sight remains constant.

In the vicinity of the neutron star, ranging from the magnetosphere to distances larger by about two orders of magnitude but still small on the system scale \citep[][Fig~2]{Kretschmar_2021a}, intrinsic variations in X-ray ionising emissions influence the structure of the stellar wind. The regions of the wind that are highly ionised, particularly those near the neutron star and in regions of low density, permit X-rays to pass through without affecting the observed absorption column density. Thus, even with no change in the amount of material along the line of sight, these fluctuations in the wind's ionisation state can lead to observed variation in the absorbing column density. 

\subsubsection{Stochastic variability inferred from MAXI long-term average profile}
\label{subsubsection:stochastic_variability_from_the_average_behavior}

The analysis of the mean profiles of both flux and hardness ratio together with their coefficient of variation (Sect.~\ref{subsection:description_flux_and_hardness_average_evolution}) reveals the stochastic nature of the system. 

\begin{figure} 
\centering
\begin{subfigure}{0.49\textwidth}
\centering
\includegraphics[width=1.0\textwidth]{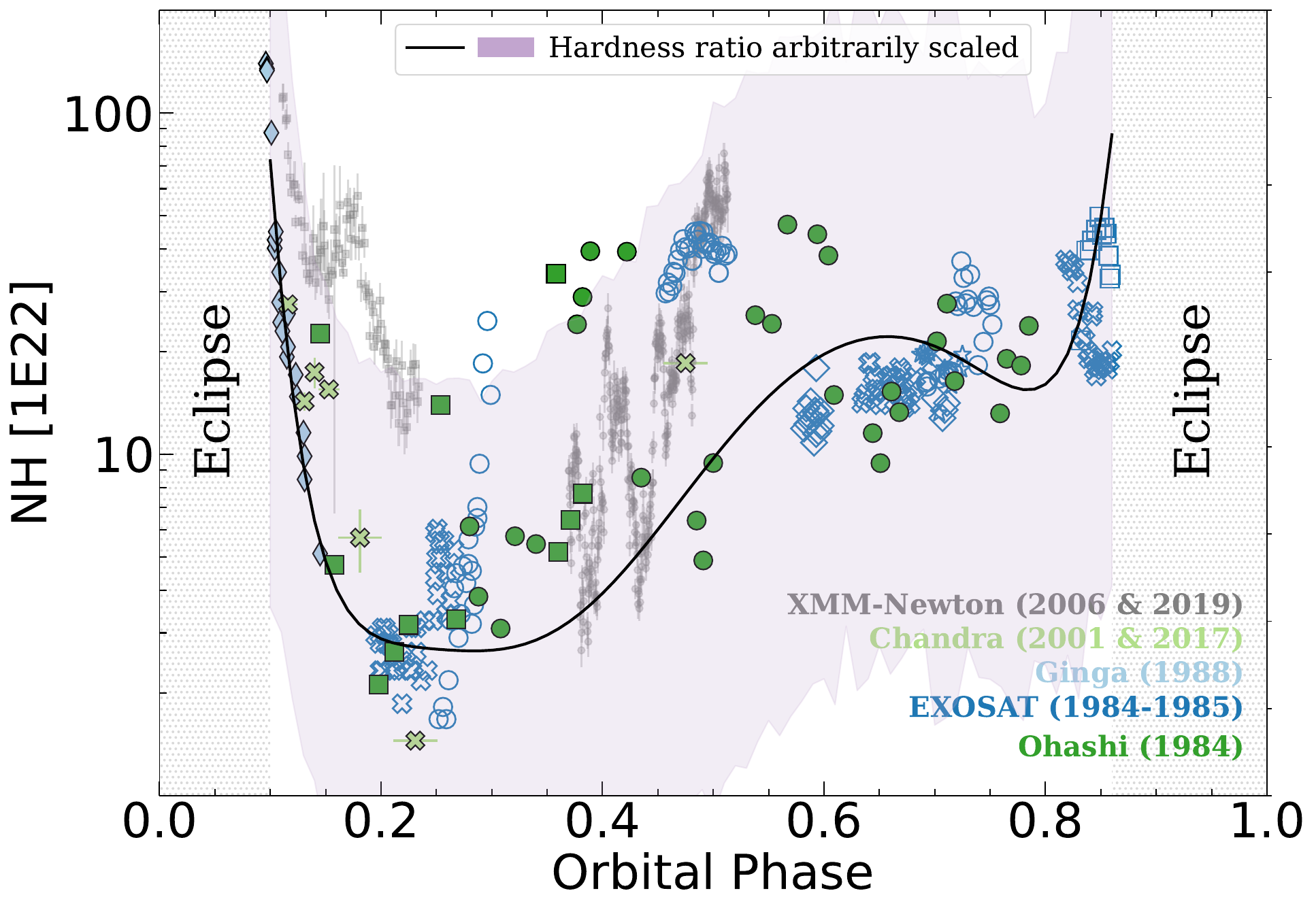}
\end{subfigure}
\caption{
Absorption column density values across the orbital phase derived from multiple X-ray satellite observations adapted from \citet{Kretschmar_2021a}. 
We refer to the accompanying text for considerations when comparing absolute $N_\mathrm{H}$ values. For a more detailed discussion on the intrinsic uncertainties associated with absolute $N_\mathrm{H}$ values, we refer to their publication.
Data taken in different binary orbits by the same mission use different symbols. 
The mean hardness profile from our MAXI/GSC sample is visually scaled and superimposed. 
The hardness ratio correlates positively with $N_{\rm H}$ but lacks a direct one-to-one relationship, hence the necessity for an arbitrary scaling. 
}
\label{figure:nh}
\end{figure}

The bump observed in the coefficient of variation of the hardness ratio (Fig.~\ref{figure:cv}c) between orbital phases 0.2 and 0.4, with an increase of approximately 50$\%$, coincides with the absence of the wake structure in the line of sight. During this phase, irregular absorption variability caused by stellar clumps becomes more significant, leading to a peak in variability. This is consistent with historically reported absorption column density values (Fig.~\ref{figure:nh}), which show greater scatter in $N_\mathrm{H}$ values during the first half of the orbit.

Following mid-orbit, a major wake structure consistently obstructs the line of sight, regardless of individual variations (Sect.~\ref{subsection:large_scale_steady_structures}). During these orbital phases, the absorption variability derived from the average trends is reduced, as the individual contributions from different binary orbits are smoothed out by the wake structure, which serves as the primary contributor to absorption.

The rapid decrease in the hardness average trend observed during eclipse egress (Fig.~\ref{figure:trends}f) is largely attributed to the influence of the extended stellar atmosphere (Sect.~\ref{subsection:large_scale_steady_structures}) and shows low variability (Fig.~\ref{figure:cv}c). However, previous studies \citep{Puls_2006, Cohen_2011, Torrejon_2015} have identified significant clumping near the photosphere of OB stars. The impact of this clumping on absorption changes during eclipse egress remains uncertain, as such changes coincide with the rapid movement of the line of sight away from the star.  

\subsubsection{Stochastic variability inferred from MAXI binary orbits}
\label{subsection:absorption_variability_in_orbit_to_orbit_hardness_profiles}

Our three-category classification (Sect.~\ref{subsection:streamlined_classification}) enables a systematic assessment of absorption variability at the level of individual binary orbits. However, the dynamics of the system necessitate a more sophisticated statistical approach to quantify the observed fluctuations (Appendix~\ref{appendix:comprehensive_variability_quantification}).  

Vela X-1 displays a standard hardness profile in $\sim$40$\%$ of cases (Fig.~\ref{figure:histogram_median}). However, we find a considerable number of hardness profiles in this category that deviate from the average hardness evolution but are within the $\pm20\%$ of the average trend. Thus, less than 40$\%$ of pointed X-ray observations will encounter conditions that deviate significantly from the average orbit. We refer to \citet[][Fig. A.1. and Table A.1.]{Kretschmar_2021a} for a discussion of X-ray observations of Vela X-1 reported in the literature. 

The remainder of our sample displays either softer hardness profiles with consistently lower absorption levels throughout the orbital period or harder profiles characterised by higher absorption levels (Fig.~\ref{figure:histogram_median}). 
These absorption profiles are influenced by factors such as the clumpy nature of the stellar wind and inherent variability of the source. 
Additionally, inspection of individual binary orbits reveals increased absorption occurring at earlier orbital phases before 0.4, when the wake structure crosses the line of sight. The onset of the wake structure does not consistently occur at the exact same orbital phase across different binary orbits, but fluctuates around $\sim$0.4 due to wake structure turbulence. This variability explains the smooth increase observed in the long-term average trend of the hardness ratio (Fig.~\ref{figure:trends}f) towards higher absorption values during the latter half of the orbit (cf. Sect~\ref{subsubsection:stochastic_variability_from_the_average_behavior}). 

Overall, the hardness profiles exhibit variability from one orbit to the next, and no periodicities are found in the 14 years of MAXI observations.

\subsection{The link between spin variations and X-ray absorption}
\label{subsection:spin_state_connection_discussion}

The continuous monitoring of the neutron star's spin period over the past decades has revealed discernible phases characterised by steady increases and decreases in rotation rates over weeks to months \citep{Malacaria_2020}. However, these fluctuations have exhibited an overall negligible net impact over the past four decades when compared to local deviations \citep{Fuerst_2010a}. In contrast to glitches observed in radio pulsars, the alterations in spin periods of neutron stars in X-ray binaries are believed to arise from accretion-induced torques, in particular in Vela X-1 where the stellar wind captured by the neutron star has enough angular momentum to circularise before it reaches the magnetosphere \citep{El_Mellah_2019a}. This hypothesis implies a dependence on the mass accretion rate, specifically on the quantity of plasma surrounding the neutron star and capable of absorbing X-ray emissions from its magnetic poles. Consequently, spin up periods would be characterised by higher X-ray hardness, i.e. periods of higher accretion. 

We investigate this hypothesis by combining our sample with the time series of the neutron star spin period from \citet[][Fig. 1]{Liao_2022}, who examine spectral properties, column densities, and Swift/BAT flux orbital profiles during spin-up and spin-down episodes. We select MAXI binary orbits from our sample (Sect.~\ref{subsection:binary_orbit_selection}) based on their spin-up/spin-down intervals, resulting in 44 binary orbits classified as spin-up and 37 as spin-down. 

The mean of the 1-day binned hardness values from the selected binary orbits is calculated separately for spin-up and spin-down states, resulting in $0.453\pm0.012$ and $0.384 \pm 0.013$, respectively. 
We use a Kolmogorov-Smirnov test to confirm that the difference in mean hardness values between the two spin groups is statistically significant. 
According to our criteria to quantify hardness ratio variability (Sect.\ref{subsection:hardness_profile_classification}), the prevalence of harder-than-average hardness profiles in the spin-up group is higher compared to the spin-down group by a factor of $\sim$5. We ascribe the hardness increases to enhanced X-ray absorption along the line-of-sight, meaning that the neutron star is engulfed in a higher amount of absorbing material when it spins-up.
This is despite the higher X-ray ionising flux from the neutron star during spin-up events, for which \citet{Liao_2022} found that the 15-50~keV Swift/BAT flux was on average $\sim$$1.6$ times higher than during spin-down events. The overall increase in hardness indicates that most of the absorbing material will be at distances from the neutron star where the ionisation parameter is too low to have a significant impact, in line with the predicted large structures.
However, there is substantial scatter in both groups: some spin-up (spin-down) episodes are associated with softer-than-average (harder-than-average) emission, and when examining hardness evolutions at the level of individual binary orbits, no clear distinction is apparent.

\citet{Liao_2022} suggests that spin-up (spin-down) could be induced by a prograde (retrograde) disk, with disk rotation switches controlled by stellar wind variations over tens of days. However, line-driven winds from massive stars do not show periodic variations on these time scales, and it is unclear whether enhanced stellar mass loss is due to changes in wind speed and/or density. While wind density changes would not affect the specific angular momentum of the accreted flow, slower winds lead to a prograde disk, while faster winds lower the circularization radius below the magnetosphere radius, resulting in quasi-spherical accretion \citep{El_Mellah+Casse_2017}. Stochastic variations could still temporarily form a prograde or retrograde disk, but clumps are too small to match the required scales \citep{Sundqvist_2018a}.

Spinning-down accretion-induced torques do not necessarily require retrograde disks: they can be achieved either through a magneto-centrifugal gating mechanism commonly called the propeller regime \citep{Bozzo_2008}, or through quasi-spherical subsonic accretion \citep{Shakura_2012}. The propeller regime is triggered when the magnetosphere radius (which is close to the inner edge of the disk) is larger than the co-rotation radius. The plasma at the inner edge of the accretion disk has to spin up in order to couple to the dipolar magnetic field lines in solid rotation with the neutron star. Therefore, the plasma experiences an additional centrifugal force which can lead to its ejection, or at least halt its penetration into the neutron star magnetosphere. Alternatively, in the quasi-spherical subsonic accretion model, the plasma does not have enough angular momentum to form a disk. Instead, it settles down on the neutron star magnetosphere as an extended quasi-static shell. The accretion into the magnetosphere is mediated by Compton cooling and convective motions in the shell.

Our interpretation for the observed episodes of spin-up and down is based on these two models. The serendipitous capture of a wind clump or a temporary increase in the overall stellar wind density could trigger transitions between the propeller regime ---where the neutron star is spun down as plasma accumulates at the outer rim of its magnetosphere--- and a regime of direct accretion, where accretion-induced torques spin-up the neutron star. Yet, the absence of correlation between the sign of the accretion-induced torque and the X-ray flux indicates that the propeller is not enough to explain the behaviour of the neutron star spin. Instead, we speculate that the capture of a clump can significantly lower the specific angular momentum of the flow and temporarily bring the system in the quasi-spherical subsonic configuration. In this case, positive and negative torques can be produced independently of the X-ray flux. More importantly, the spin-up episode can be sustained as long as the reservoir of plasma which was present at the outer rim of the neutron star magnetosphere contains enough material. Therefore, it lasts much longer (typically a few weeks) than the stellar wind variation which triggered it, be it clump capture or enhanced wind density. This interpretation accounts for the predominance of higher X-ray hardness ratios during the spin-up episodes. Also, it would explain why we observe such a huge scatter around the aforementioned hardness ratios mean values: even when wind density at the orbital scale comes back to its standard values, the spin-up episodes continue. 

\section{Conclusion and outlook}
\label{section:conclusion}

In this study, we conduct a comprehensive analysis of the long-term evolution and variability of X-ray absorption in the wind-accreting HMXB Vela X-1 using all-sky monitor MAXI data. We demonstrate that $N_\mathrm{H}$ is the primary factor influencing changes in hardness ratio and utilize the latter as a tracer of absorption variability. This technique enables the analysis of absorption-induced variations across the entire orbital period. 

The novelty of this work lies in the investigation of absorption variability within individual binary orbits over 14 years of continuous observations. Our principal findings are: 
\begin{itemize}
    \item Large-scale structures, such as the accretion wake trailing the neutron star, significantly influence the observed absorption patterns, consistent with existing literature on the impact of hydrodynamic interactions between the neutron star and the stellar wind. 
    \item There is significant variability in hardness profiles from one orbit to the next, without any discernible periodic pattern. This variability underscores the heterogeneous composition of wind structures. The stochastic variability of absorption, mainly attributed to clumps in the stellar wind, contributes to short-term absorption variations. 
    \item Less than 40$\%$ of the hardness profiles follow a typical hardness evolution within a 20$\%$ difference from the average trend. Therefore, pointed X-ray observations are likely to encounter non-average scenarios, explaining the diversity of $N_{\rm H}$ values observed at the same orbital phase in the literature. 
    \item Spin-up events in Vela X-1 correlate positively with harder-than-average binary orbits, reflecting increased absorption along the line of sight, which indicates a higher density of surrounding absorbing material. The observed spin variations are attributed to accretion-induced torques, with spin-up episodes sustained by a reservoir of plasma around the neutron star's magnetosphere. This suggests an interplay between the accreted plasma, variations in the stellar wind from the donor star, and the neutron star's magnetosphere.
\end{itemize}

The statistical approach followed in this study demonstrates its applicability to other similar sources monitored by MAXI, suggesting its potential for broader usage in understanding absorption variability in binary systems. Our findings motivate new considerations in theoretical studies, urging an increase in the number of modelled orbits. Future theoretical investigations should account for these factors to better capture the observed absorption dynamics in binary systems.

Moreover, the enhanced sensitivity of the upcoming Einstein Probe mission \citep{Weimin_2022} holds promise for conducting similar studies with greater detail. Its improved capabilities compared to current observatories will enable deeper exploration into the complexities of X-ray absorption variability, providing further insights into the dynamics of stellar winds and accretion processes onto neutron stars. 

\begin{acknowledgements} 
We thank the anonymous referee, who has improved the clarity of this manuscript.
LA acknowledges the support by cosine measurement systems, and  
SMN acknowledges funding under project PID2021-122955OB-C41 funded by MCIN/AEI/10.13039/501100011033 and by 'ERDF A way of making Europe'. 
We acknowledge support from ESA through the Science Faculty - Funding reference ESA-SCI-SC-LE-181, the Institute of Space Science (ICE) - CSIC, and the Institut d'Estudis Espacials de Catalunya (IEEC). 
This research was supported by the International Space Science Institute (ISSI) in Bern, through ISSI International Team project \#495 (Feeding the spinning top). 
We thank Zhenxuan Liao's team at Sanming University in China for sharing the time intervals in the spin frequency history used in their research. \\
This research has made use of 
(1) the MAXI data provided by RIKEN, JAXA, and the MAXI team; 
(2) the Interactive Spectral Interpretation System (ISIS) \citep{Houck_2000, Houck_2002, Nobel_2008} maintained by the Chandra X-ray Center group at MIT; 
(3) the ISIS function (\texttt{isisscripts})\footnote{\url{http://www.sternwarte.uni-erlangen.de/isis/}} provided by ECAP/Remeis observatory and MIT; 
(4) NASA's Astrophysics Data System Bibliographic Service (ADS);
(5) the Astropy package\footnote{\url{https://www.astropy.org/}} \citep{astropy_collab_2013, astropy_collab_2018} and Matplotlib library\footnote{\url{https://matplotlib.org/}} \citep{Hunter_2007} for data analysis and visualisation 
\end{acknowledgements}

\bibliographystyle{aa}
\bibliography{VelaX1_MAXI}

\begin{appendix}
    
\section{Detailed statistical analysis of hardness ratio variability in binary orbits}
\label{appendix:comprehensive_variability_quantification}

The method described in Sect.~\ref{subsection:streamlined_classification}, based on the $F_\mathrm{50}$ parameter (Eq.~\ref{equation:F_50}), provides a general overview sufficient for drawing the main conclusions of this study on the system's long-term evolution using MAXI data. However, Fig.~\ref{figure:orbits}d,e,f reveals irregularities within each defined group that warrant further quantification. In this appendix, we detail a more rigorous statistical method to quantify the variability of hardness ratios with respect to the average trend in each binary orbit. This approach allows for a comprehensive description of the evolution of individual hardness profiles in the MAXI dataset without the need for manual inspection. 

\subsection{Defining two new statistics}
\label{appendix:defining_two_new_statistics}

After dividing our sample based on the $F_\mathrm{50}$ parameter (Sect.~\ref{subsection:streamlined_classification}), we employ two new statistics to explore the dispersion observed among the hardness ratios within each category: $\chi^2$ (Eq.~\ref{equation:chi2}), and $D_\mathrm{orb}$ (Eq.~\ref{equation:d}). While profiles with higher dispersion could be categorised into a fourth group of irregular profiles, doing so would necessitate defining additional criteria for each of the new statistics, adding unnecessary complexity to our approach. Instead, we identify hardness evolutions that deviate from the standard, harder, and softer profiles without grouping them into a new category, yet highlighting the intrinsic orbit-to-orbit variability. 

The standard hardness profiles are expected to align with the average hardness trend. To quantify the degree of dispersion of the hardness ratios within this profile group, we employ the $\chi^2$ metric, defined as follows: 
\begin{equation} \label{equation:chi2}
\chi^{2} = \frac{1}{N} \sum_{i=1}^{N} \left(\frac{x_{i} - E_{i}}{\sigma_{i}}\right)^{2},
\end{equation}
where $N$ is the number of data points in the binary orbit, $x_{i}$ is the 1-day binned hardness value, $E_{i}$ is the expected value based on the average trend, and $\sigma_{i}$ is the uncertainty associated with $x_{i}$. 

The $\chi^2$ metric is solely employed to quantify deviations in standard profiles, as harder and softer profiles exhibit high $\chi^2$ values by definition. For standard orbits, low $\chi^2$ values correspond to orbits characterised by a $F_\mathrm{50}$ parameter close to 0.5, indicating visual alignment with the hardness average trend. However, high $\chi^2$ values may stem from two distinct scenarios. Firstly, orbits where the majority of hardness ratios deviate either above or below the hardness average trend, but consistently in the same direction. Consequently, the associated $F_\mathrm{50}$ parameter tends towards the extreme values of the defined interval for standard profiles, i.e. 0.4 < $F_\mathrm{50}$ < 0.6. Secondly, orbits exhibiting equal deviations either above and below the average trend, resulting in $F_\mathrm{50}$ being approximately 0.5. 

We define a secondary statistic to quantify the dispersion in softer and harder profiles. Beginning with the (1$-$CDF) values of each hardness ratio (Sect.~\ref{subsection:quantifying_variability_with_CDF}), we calculate the mean of these values within each orbit to establish the parameter $F_\mathrm{av}$, as follows:
\begin{equation} \label{equation:F_av}
F_\mathrm{av} = \mathrm{mean}\sum_{i=1}^{N} (1-CDF)_i,
\end{equation}
where $N$ is the number of hardness ratios in the binary orbit, and (1$-$CDF) is the complementary CDF associated with each hardness ratio.  This metric is particularly sensitive to outliers. 

Subsequently, we define the coefficient $D_\mathrm{orb}$:
\begin{equation} \label{equation:d}
    D_\mathrm{orb}=\frac{F_\mathrm{50}}{F_{\mathrm{av}}},
\end{equation}
where $F_\mathrm{50}$ is the median (Eq.~\ref{equation:F_50}), and $F_\mathrm{av}$ is the mean (Eq.~\ref{equation:F_av}) of the (1$-$CDF) values for the hardness ratios of each profile.

$D_\mathrm{orb}$ denotes both the magnitude and direction of the outliers. For example, $D_{\mathrm{orb}}<1$ indicates that $\mathrm{F_{av}} > \mathrm{F}_{50}$, i.e. the presence of outliers towards harder hardness ratios. Conversely, $D_{\mathrm{orb}}>1$ denotes outliers towards softer hardness ratios, while a $D_\mathrm{orb}$ around 1 indicates no dispersion. However, as mentioned above, in standard profiles with an even number of 1-day binned hardness ratios and high $\chi^2$ values with $F_\mathrm{50}$ around 0.5, we would observe a dispersion around the mean pattern that $D_\mathrm{orb}$ would not be able to discern, as the coefficient would also be around 1. Therefore, we restrict the use of this coefficient exclusively to softer and harder profiles. 

\subsection{Visual summary of orbit-to-orbit variability}
\label{appendix:visual_summary_of_orbit_to_orbit_variability}

\begin{figure} 
\centering
\includegraphics[width=0.49\textwidth]{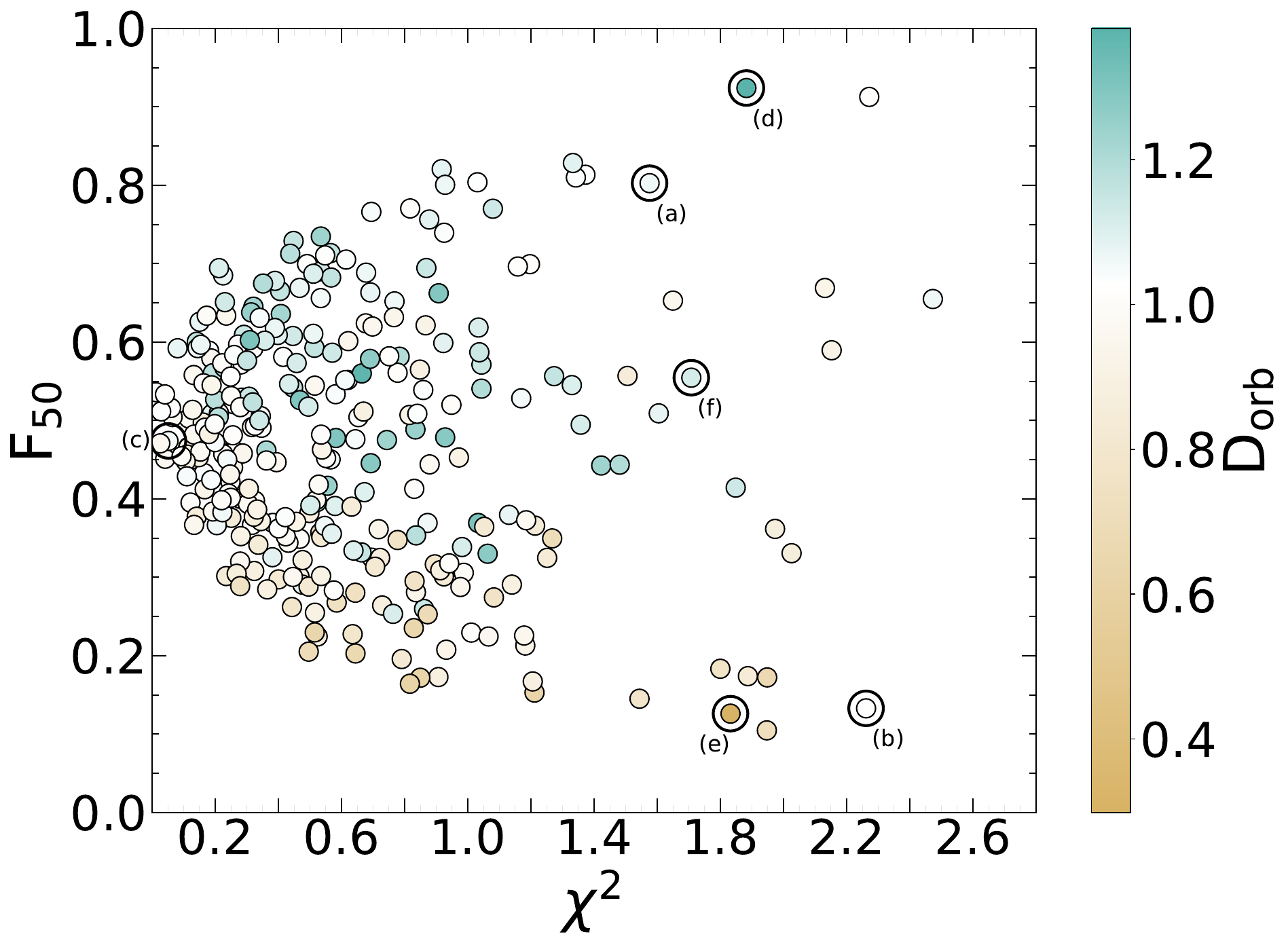}
\caption{
Orbit-to-orbit hardness variability relative to the hardness average trend. Each point represents an individual binary orbit from our sample (Sect.~\ref{subsection:binary_orbit_selection}), with $F_\mathrm{50}$ (Eq.~\ref{equation:F_50}) values plotted against $\chi^2$ (Eq.~\ref{equation:chi2}). $D_\mathrm{orb}$ coefficient (Eq.~\ref{equation:d}) is colour-coded. We highlight the binary orbits shown in Fig.~\ref{figure:orbits} with a black circle. Further details in the text.
}
\label{figure:binary_orbit_map}
\end{figure}

With these three parameters ($F_\mathrm{50}$, $\chi^2$, and $D_\mathrm{orb}$), we assess the orbit-to-orbit hardness variability. In Fig.~\ref{figure:binary_orbit_map}, we locate each orbit of our sample in a $\chi^2$-$F_{50}$ space and colour-code them according to $D_\mathrm{orb}$. The hardest hardness profiles are located at the top of the diagram, descending towards softer hardness profiles at the bottom. The extreme $F_\mathrm{50}$ values associated with high $\chi^2$ values (harder and softer profiles), and the central $F_\mathrm{50}$ values associated with low $\chi^2$ values (standard profiles), result in the distinctive C-shaped pattern. 

For instance, the harder (Fig.~\ref{figure:orbits}a), softer (Fig.~\ref{figure:orbits}b), and standard (Fig.~\ref{figure:orbits}c) orbits represent profiles without deviation, hence marked with a colour code associated with D$_{\mathrm{orb}}\sim1$. They are positioned along the C shape in Fig. 9 because there is a visual agreement between their hardness evolution and the category to which they belong. However, the harder (Fig.~\ref{figure:orbits}d) and softer (Fig.~\ref{figure:orbits}e) orbits, although also situated along the C shape, their greater deviation assigns them a colour code indicating the presence of outliers towards softer and harder hardness values, respectively. The standard orbit (Fig.~\ref{figure:orbits}f) displays a visually more irregular hardness evolution around the average trend. It is positioned outside the C shape in Fig.~\ref{figure:binary_orbit_map}, with a colour code indicating a deviation towards harder hardness values, as we can verify when looking at the harder hardness ratios between orbital phases 0.2--0.4 of the orbit.  

The large population of orbits in Fig.~\ref{figure:binary_orbit_map} around $F_\mathrm{50}$ with increasing $\chi^2$ values, as well as the significant number of orbits indicating the presence of outliers through their $D_\mathrm{orb}$ parameter, denote the considerable variability of hardness evolution patterns in our sample, and consequently, the challenging task of identifying evolution patterns at the individual orbit level. The analysis of $D_\mathrm{orb}$ collectively for the three defined categories reveals that 55.2$\%$ of the hardness profiles are situated within $\pm$10$\%$ of the central value of $D_{\mathrm{orb}}=1$, while the rest deviates.  

\section{Light curve for individual binary orbit}
\label{appendix:light-curve-for-individual-binary-orbit}

\begin{figure} 
\centering
\includegraphics[width=0.49\textwidth]{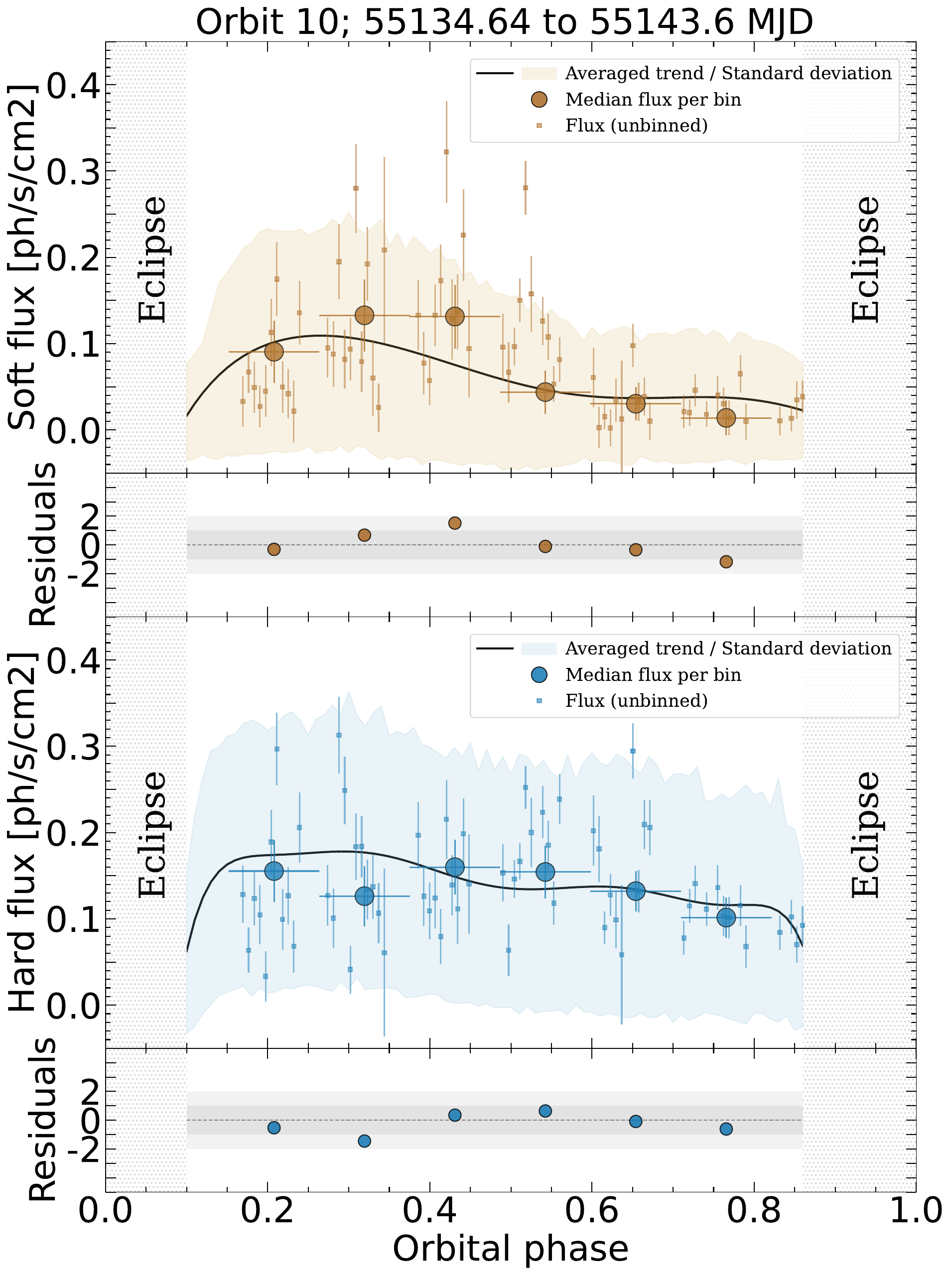}
\caption{
MAXI/GSC light curve for both the soft (top, yellow points) and hard (bottom, blue points) energy bands. Small square markers depict unbinned flux data, while large circular markers represent median flux values per phase bin, each with their respective uncertainties. The bin size is indicated by the horizontal error bars. The figure also shows the average trend as a solid black line and the standard deviation as a shaded area, consistent with Fig.~\ref{figure:trends}. 
}
\label{figure:lightcurve}
\end{figure}

Figure~\ref{figure:lightcurve} shows a example of a MAXI/GSC light curve of an individual binary orbit for both soft and hard energy bands, illustrating unbinned and 1-day binned flux data distribution and variability across orbital phase bins. 

\end{appendix}

\end{document}